\documentclass[sigconf]{acmart}
\usepackage[ruled,linesnumbered]{algorithm2e}
\usepackage{bm}
\usepackage{algorithmicx}  
\usepackage{algpseudocode}
\usepackage[normalem]{ulem}
\useunder{\uline}{\ul}{}
\usepackage{graphicx} 
\usepackage{float} 
\usepackage{subfigure} 
\usepackage{endnotes}

\AtBeginDocument{%
  \providecommand\BibTeX{{%
    \normalfont B\kern-0.5em{\scshape i\kern-0.25em b}\kern-0.8em\TeX}}}

\setcopyright{acmcopyright}
\copyrightyear{2024}
\acmYear{2024}
\acmDOI{XXXXXXX.XXXXXXX}

\acmConference[Conference acronym 'XX]{Make sure to enter the correct
  conference title from your rights confirmation emai}{June 03--05,
  2018}{Woodstock, NY}
\acmPrice{15.00}
\acmISBN{978-1-4503-XXXX-X/18/06}

\begin{document}

\title{Pareto-based Multi-Objective Recommender System with Forgetting Curve}

\author{Jipeng Jin$^1$, Zhaoxiang Zhang$^1$, Zhiheng Li$^1$, Xiaofeng Gao$^{1,*}$, Xiongwen Yang$^2$, Lei Xiao$^2$, Jie Jiang$^2$}
\affiliation{%
  \institution{$^1$Shanghai Jiao Tong University \hspace{0.3em} $^2$Tencent}
  \institution{\{jinjipeng, zzx-sjtu, 2015984620,gaoxiaofeng\}@sjtu.edu.cn, \{captainyang, shawnxiao, zeus\}@tencent.com}
  \country{China}
}

\renewcommand{\shortauthors}{Jipeng Jin, ZhaoXiang Zhang and Zhiheng Li, et al.}
\renewcommand\footnotetextcopyrightpermission[1]{}
\begin{abstract}

Recommender systems with cascading architecture play an increasingly significant role in online recommendation platforms, where the approach to dealing with negative feedback is a vital issue. 
For instance, in short video platforms, users tend to quickly slip away from candidates that they feel aversive, and recommender systems are expected to receive these explicit negative feedbacks and make adjustments to avoid these recommendations.
Considering recency effect in memories, we propose a forgetting model based on Ebbinghaus Forgetting Curve to cope with negative feedback. 
In addition, we introduce a Pareto optimization solver to guarantee a better trade-off between recency and model performance.
In conclusion, we propose Pareto-based Multi-Objective Recommender System with forgetting curve (PMORS), which can be applied to any multi-objective recommendation and show sufficiently superiority when facing explicit negative feedback.
We have conducted evaluations of PMORS and achieved favorable outcomes in short-video scenarios on both public dataset and industrial dataset. After being deployed on an online short video platform named WeChat Channels in May, 2023, PMORS has not only demonstrated promising results for both consistency and recency but also achieved an improvement of up to +1.45\% GMV.

\end{abstract}



\maketitle
\section{Introduction}

\begin{figure}[ht]
  \centering
  \setlength{\abovecaptionskip}{0pt}
  \includegraphics[width=\linewidth]{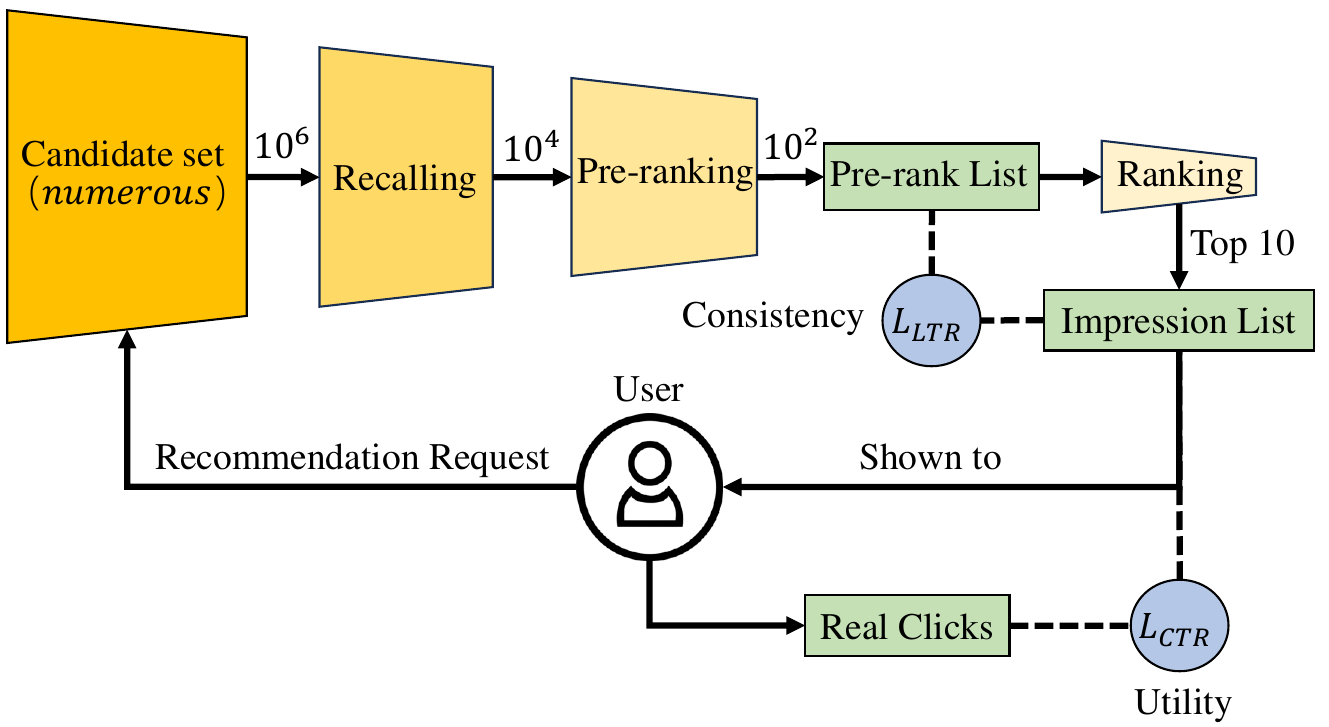}
  \caption{The Process of A Typical Recommender System }
  \label{RS process}
  \Description{}
\end{figure}

Recommender systems have gained widespread adoption and have become an integral part of various fields in industry, including e-commerce, streaming platforms, social media, and content recommendation platforms. 
Recommender systems have proven to be highly effective in enhancing user experience, increasing user engagement and driving business revenue~\cite{RS2022survey}. 
Due to the strict time performance requirements of online recommender systems, a commonly employed solution in practical industrial systems is the adoption of a cascading architecture~\cite{2023slate-aware,2023CT4Rec,2023integrated}, wherein a lightweight pre-ranking stage is conducted before recommending candidates to users through the ranking stage. 
Pre-ranking stage often employs simpler and faster models compared to the ranking stage, aiming to reduce the candidate pool and enable low-latency recommendation for subsequent ranking stage.
The process of a typical recommender system, initiated by a user's recommendation request, is illustrated in Fig.~\ref{RS process}. 
Recommender models in the ranking stage prioritize enhancing the utility objective like Click-Through Rate (CTR) or ConVersion Rate (CVR). 
While models in the pre-ranking stage focus on the \textit{consistency} between pre-ranking and ranking stage~\cite{2023berd+,2023device,2023relationship}. 

\vspace{-10pt}

In the practical application of recommender systems in industries, negative feedback from users is commonly encountered and these negative feedback explicitly manifests users' aversion towards some certain recommendations~\cite{2023negative,2023revisiting,2023learning}.
On short video platform, users tend to quickly slip away from videos that they dislike. Each fast-slip action by the user represents explicit aversion towards the exposed candidate, which should be imposed penalty for better recommendation.
If these negative feedback can not be effectively addressed, this can lead to a decline in user experience and even result in user attrition. What's worse, recommender systems may make erroneous personalized preference judgments for users, thereby impacting the performance and accuracy of recommendations~\cite{2019drcgr,2022self}. 
Wu et al.~\cite{wu2020neural} computed a score for negative feedback to assist the recommender system in learning and improving, while Wang et al.~\cite{wang2021interactive} applied an encoder and decoder to deal with feedback. Diverging from conventional methods, our approach places a stronger emphasis on user experience by employing the forgetting curve to model negative feedback from users.

There exists \textit{Recency Effect} in memories\cite{faggioli2020recency,nitu2021improvising}, which means that recently encountered information predominates the overall impression, exerting a greater influence compared to the previously obtained information.
Regarding the specific context of recommender systems, users passively reinforce their memory and form new positive/negative memories with each exposure to a candidate. Each exposure to the candidate serves as a \textit{review} of the impression that the candidate leaves to the user. To effectively measure the impact of negative feedback, it is crucial to establish a quantifiable metric that can accurately assess the influence of aversive memory. 
Therefore, we introduce a new metric, namely FastSlipRate specifically for short video platform, to quantify the impact on subsequent recommendations and develop corresponding penalty strategies, which is illuminated and supported by Ebbinghaus Forgetting Curve. 
Ebbinghaus Forgetting Curve~\cite{1885memory,2014modeling} is used to describe the forgetting rate of mid-term memory, which explains that memory strength decreases as time elapses. 
Nevertheless, both offline experiments and online applications demonstrate that independently optimization of negative feedback will, instead, have an adverse impact on the overall performance of the recommender system. Excessive punishment can introduce additional noise, thereby interfering with the learning of recommendation metrics.

Hence, it is imperative to employ multi-objective recommendation approach that optimizes negative feedback without compromising the utility and consistency of the recommender system. 
For most existing multi-objective recommender systems, it is common practice to assign different weights to different objectives and then transform these multiple objectives into a single objective function for the purpose of the optimal solution~\cite{2022toward,2022review,2023tourism}.
However, determining the weights for different objectives is often a challenging task, as it is difficult to achieve a well-balanced and satisfactory trade-off among the diverse objectives. Unlike commonly used methods of specifying hyperparameters~\cite{20231hyperparameter}, we utilize the Pareto theory for better trade-offs among multiple objectives.

Pareto optimization methods are widely regarded as a highly effective approach for addressing multi-objective problems, aiming to optimize multiple objectives towards Pareto efficiency, which represents the optimal trade-off among the objectives in a given problem domain. Pareto efficiency is a state when it is impossible to improve one objective without hurting other objectives in terms of multi-objective optimization~\cite{2019Pareto}. Unlike the commonly used evolutionary algorithms~\cite{2019indexed,2021multi,2023distributional} or scalarization methods~\cite{2018modeling,2020progressive,2021selective,2021drivebfr}  employed by Pareto in multi-objective systems, we apply a \textit{Pareto Optimization Solver} to optimize multiple objectives in our paper, which is inspired by multiple-gradient descent algorithm (MGDA)~\cite{MGDA2012} and allows adaptive adjustments of weights for multiple objectives.

To this end, we proposed Pareto-based Multi-Objective Recommender System with forgetting curve (PMORS), which can be applied to any multi-objective recommendation and show sufficiently superiority when facing explicit negative feedback. 
We conduct extensive experiments on public dataset from a video platform to prove the effectiveness and scalability of our model. 
PMORS has been deployed on WeChat Channels, an online short video platform of Tencent,  and demonstrated promising results.

Our main contributions can be summarized as follows:
\vspace{-10pt}
\begin{itemize}
    \item We propose a  Pareto-based Multi-Objective Recommender System with forgetting curve (PMORS) for multi-objective recommendation with negative feedback.
    \item To the best of our knowledge, we are the first to employ forgetting curve in dealing with negative feedback, which provides a more reasonable and efficient approach.
    \item We employ Pareto theory to achieve a better trade-off among multiple objectives when imposing penalty on negative feedback guided by forgetting curve. 
    \item We conduct extensive experiments on public dataset and deploy our model in WeChat Channels. In both scenarios, PMORS has demonstrated promising results for both \textit{consistency} and \textit{recency}. And it also achieves an improvement of up to +1.45\% GMV in online environment.
\end{itemize}
\vspace{-10pt}

\section{Related Works}
\textbf{Pareto Optimization in Recommender Systems.} Some studies attempt to model the trade-offs among mutiple objectives in recommender systems~\cite{2023copr,CONCHACARRASCO2023,ISUFI2021}, where Pareto optimization is expected to make a critical difference. 
Pareto optimal solution is a state when it is impossible to improve one objective without hurting other objectives in terms of multi-objective optimization~\cite{2019Pareto}. 
Recently, it is pointed out that the Karush-Kuhn-Tucker (KKT) conditions can be used to guide Pareto optimization and Multiple-Gradient Descent Algorithm (MGDA) has been applied to find a Pareto optimal solution~\cite{MGDA2012}. 
PE-LTR~\cite{2019Pareto} was the first to provide a Pareto efficient framework for multi-objective recommendation with theoretical guarantees, where multiple objectives may conflict with each other. 
PAPERec~\cite{2021paretoQM} utilized reinforcement learning to find a Pareto optimal solution with personalized preference weights.
Moreover, some recent research~\cite{2022multistakeholderPareto,2021inded} proposed superior models based on Pareto theory for recommender systems, better balancing the objectives trade-offs. 
Meanwhile, the theory of Pareto optimization is constantly developing and innovating, and some novel methods or models have been proposed to achieve Pareto optimal solution with more efficiency and better performance~\cite{2020continuousPareto,2023post-hoc}.

\noindent \textbf{Multi-Objective Recommender System.} As stated in~\cite{2023multi-objectivesurvey}, while traditional recommender systems often focus on a single objective, Multi-Objective Recommender Systems(MORS) aims to balance diverse conflicting objectives such as accuracy, novelty, and diversity to offer more various and customized recommendations to users. Based on the different choice of objectives, typical applications of multi-objective optimization in recommender systems involve RS with Multiple Quality Metrics\cite{2021paretoQM,2021MOQM},Group RS~\cite{2023enhancing,2023improving}, Multi-stakeholder RSs~\cite{2023deep,2023categorization}, Multi-task RSs~\cite{2021attribute,2023multi} and Clustering \& Rule Based RSs\cite{2019improve,2022modeling}. To optimize these subfields, a line of approaches have been applied, particularly the two widely-used, Evolutionary Algorithms(EAs)~\cite{2019indexed,2021multi,2023distributional} and Scalarization~\cite{2018modeling,2020progressive,2021selective,2021drivebfr}. EAs are based on simulating the process of natural selection in order to reach the optimal results that get different trade-offs among various objectives. 
Despite their effectiveness, EAs can be timeconsuming for its slow convergence speed and become hard to employ for its non-gradient methology. Scalarization approaches seek to discover exact solutions by mathematical models and algorithms.Existing studies mainly focus on two directions, to optimize the local structure like studies of MMoE~\cite{2018modeling},PLE~\cite{2020progressive} and to optimize the losses of various objectives~\cite{2021selective,2021drivebfr}. 
Although they have limits on a large number of objectives or constraints, they are appreciated by their simplicity, versatility, and ease of interpretation.

\noindent \textbf{Forgetting Curve.} Forgetting curve is used to describe the forgetting rate of mid-term memory, which was first expressed as an exponential formula by psychologist Hermann Ebbinghaus based on his experiments~\cite{1885memory}. It has been widely used in diverse fields such as production scheduling support, management of industrial costs and optimization of workforce~\cite{ferreiraforgetting}. Considering that the user passively strengthens their memory and forms new positive/negative memories with each exposure to the displayed candidate sequences, a few studies~\cite{2006primacy,2009consumer,2013social} suppose that people's reactions to recommendation like clicking also follow a related psychologic rule named recency effect. Therefore, some researchers~\cite{2014modeling,2021fuzzy,2022building} have applied these forgetting curves into recommender systems to estimate that impact. ~\cite{2014modeling} proposes interest forgetting curves(IFC) based on Ebbinghaus' work to evaluate the time-variant recency impact on users' interest on music recommendation.~\cite{2022building} proposed a label-based memory forgetting enhancement model (LMFE) by combing traditional forgetting curves and a simple enhancement model to improve TV recommendation. While these examples have successfully applied the curves into local sturctures of their models, they ignore its value of being an independent optimization objective cooperated with other common objectives like utility and diversity. 

\noindent \textbf{Hyperparameter Optimization.} One of the most commonly used methods for hyperparameter optimization is grid search, which discretizes the range of each hyperparameter and exhaustively evaluates every combination of values~\cite{gridsearch}. Another simple and efficient method is random search, preferable to grid search and a surprisingly strong baseline for hyperparameter optimization in many practical settings~\cite{2023hyperparameter}. Aimed to achieve better performance with efficiency, other mathematical methodologies have been applied  to the process of hyperparameter optimization, such as iterated racing~\cite{2016irace} and bayesian optimization~\cite{2022bayesianhyperparameter,2023bayesianhyperparameter}. Recently, Pareto optimization theory has come into notice and some researchers have attempted to optimize hyperparameter by Pareto theories~\cite{2022ParetoRS,2023multi-objectivesurvey}, expected to achieve better trade-offs for multiple hyperparameters.

\section{Methodology}

\subsection{Preliminary}

\subsubsection{Pareto Theory}

Given a multi-objective problem aiming to 
optimize $t$ loss functions $\mathcal{L}_1(\bm{\theta}),\dots, \mathcal{L}_t(\bm{\theta})$, where $\bm{\theta}$ is a solution to the optimization.

\vspace{-3pt}
\begin{definition}[Pareto Dominance] 
    We say that one solution $\bm{\theta}_1$ dominates another solution $\bm{\theta}_2$ if: 
    for all objectives $\mathcal{L}_i(\bm{\theta}_1) \leq \mathcal{L}_i(\bm{\theta}_2)$, 
    $i \in \{1,\dots,t\}$, 
    and there exists at least one objective $\mathcal{L}_j(\bm{\theta})$, $  j \in \{1,\dots,t\}$ 
    satisfying $\mathcal{L}_j(\bm{\theta}_1) < \mathcal{L}_j(\bm{\theta}_1)$.

\end{definition}
\vspace{-5pt}

\begin{definition}[Pareto Optimality/Pareto Efficiency]
    A solution $\bm{\theta}$ is Pareto optimal (Pareto efficient) if there exists no other solution $\bm{\theta}'$ that dominates $\bm{\theta}$. 
    The set of all Pareto optimal solutions is called the Pareto Set. 
\end{definition}
\vspace{-3pt}

On the basis of scalarization method, we can aggregate each loss fuction $\mathcal{L}_i(\bm{\theta})$ with a weight $\alpha_i$:
\vspace{-3pt}
\begin{equation}\label{loss}
    \mathcal{L}(\bm{\theta}) = \sum_{i=1}^t \alpha_i \mathcal{L}_i(\bm{\theta}).
\end{equation}

It is hard to attach conditions for Pareto Optimality directly, so we introduce the Pareto Stationary, which is a necessary condition for Pareto optimality. 
A Pareto optimality must be Pareto stationary, and the inverse holds true under mild conditions~\cite{sener2018multi}.
\vspace{-3pt}

\begin{definition}[Pareto Stationarity] \label{Pareto stationarity}
    A solution $\bm{\theta}$ is Pareto stationarity if it satisfies
   
    Karush-Kuhn-Tucker (KKT) conditions~\cite{KKT} \footnote{Considering practical application scenarios, we have omitted several unnecessary constraints.}:
    \begin{enumerate}
        \item $\sum_{i=1}^t \alpha_i \nabla_{\bm{\theta}} \mathcal{L}_i (\bm{\theta}) = 0$ ,
        \item $\sum_{i=1}^t \alpha_i = 1, \ \alpha_i \geq 0$, for $i = 1, \dots, t$.

    \end{enumerate}
   
\end{definition}

\subsubsection{Ebbinghaus Forgetting Curve}\label{Ebbinghauscurves}
Memory rate $R(t)$ refers to the proportion of memory content in relation to the original content at time $t$.
Ebbinghaus Forgetting Curve explains that memory strength decreases as time elapses, which can be represented as:
\vspace{-3pt}
\begin{equation}\label{forgettingcurve}
    R(t) = e^{-\frac{t}{S}} , 
\vspace{-3pt}
\end{equation}
where $S$ is relative memory strength, which is a constant.

There are 2 characteristics for Ebbinghaus Forgetting Curve:
\vspace{-1pt}
\begin{enumerate}
    \item Memory strength decays over time and exhibits an exponential decay.
    \item $S$ varies depending on the specific memory content. 
\end{enumerate}
\vspace{-3pt}

\subsection{Proposed Framework}

\begin{figure*}[ht]
  \centering
  \setlength{\abovecaptionskip}{0pt}
  \includegraphics[width=\linewidth]{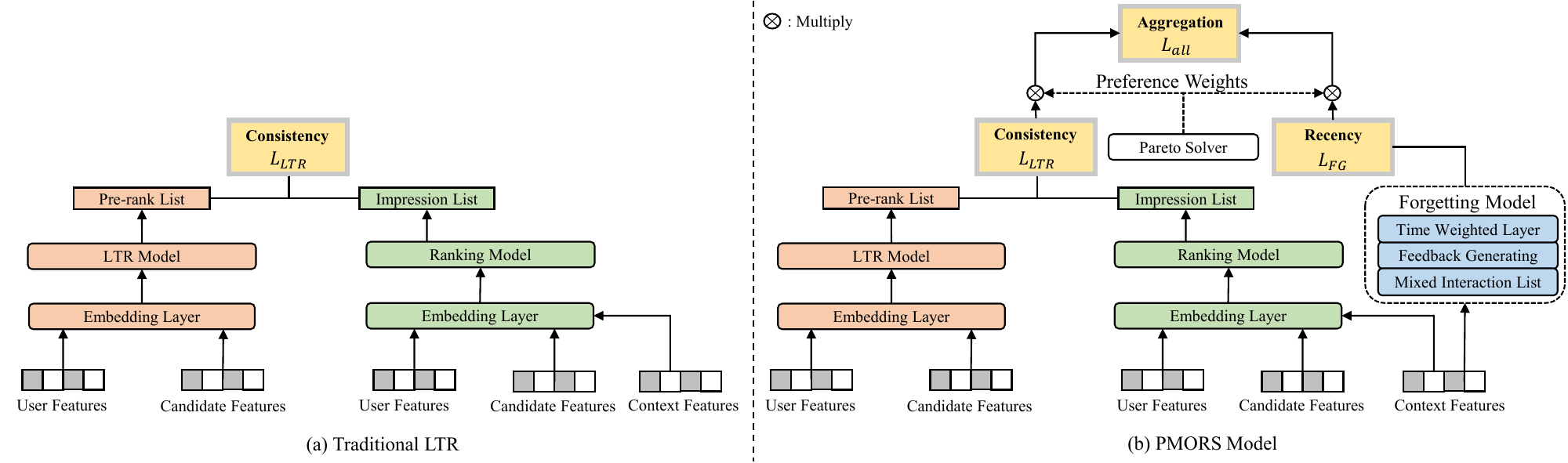}
  \caption{(a) shows the structure of traditional LTR, consisting of a pre-ranking model (orange) and a fixed pretrained ranking model (green). (b) illustrates the forgetting model (blue) which is used to estimate a new objective called \textit{recency} and a Pareto optimization solver which aggregates the two losses and purposes preference weights that lead the model to Pareto efficiency. }
  \label{modelimage}
  \Description{}
\end{figure*}

Our model consists of Pareto optimization solver and three models,i.e., forgetting model, pre-ranking model, and ranking model, whose structure is shown in Fig.\ref{modelimage}. Note that our model is based on the industrial practice of pre-ranking model of WeChat Channels. Hence, the proposed model framework is built upon pre-ranking and can be generalized to other stages as well.

\subsubsection{Ranking and Pre-ranking Model}
\label{ranking and pre-ranking}
Ranking model is a fundamental component in recommender systems, as it determines the order in which the candidates are presented to users based on the ranking scores. 
During the ranking stage, more sophisticated algorithms and models are employed to sort the candidates, taking various factors into account such as user's historical behavior, interest and preferences. 
The intricate details of the ranking model's operation are of secondary importance, as its primary function is to generate ranking scores for the candidates,
which are employed in the subsequent pre-ranking model for training.

Pre-ranking model is the preliminary sorting process of candidates before the ranking model is applied to generate final recommendations. It is designed to filter the candidate set and select most relevant items for the following ranking stage. 
Pre-ranking model adopts learning-to-rank method to model each candidate’s accurate rank in the impression list generated by ranking models, and then produces a pre-rank list of candidates, which concentrates more on the ranking instead of scores.
The difference between two lists will be calculated by $L_{LTR}$:
\begin{equation}\label{preranking}
    L_{LTR} = 
    \begin{cases}
    \sum\limits_{i,j}  \log \Big(1+ e^{-\big(p(\mathbf{x}_i)-p(\mathbf{x}_j)\big)}\Big), &q(\mathbf{x}_i) >q(\mathbf{x}_j)  \\
    0, &q(\mathbf{x}_i) \leq q(\mathbf{x}_j) 
    \end{cases},
\end{equation}
where $\mathbf{x}_i$, $\mathbf{x}_j$ denotes the input features of models, $p(\mathbf{x}_i)$, $p(\mathbf{x}_j)$ is the scores of pre-ranking model, and $q(\mathbf{x}_i)$, $q(\mathbf{x}_j)$ is the scores of ranking model, respectively.

\subsubsection{Forgetting Model}\label{FORGETTING MODEL}

\begin{figure}[htbp]
  \centering
  \includegraphics[width=\linewidth]{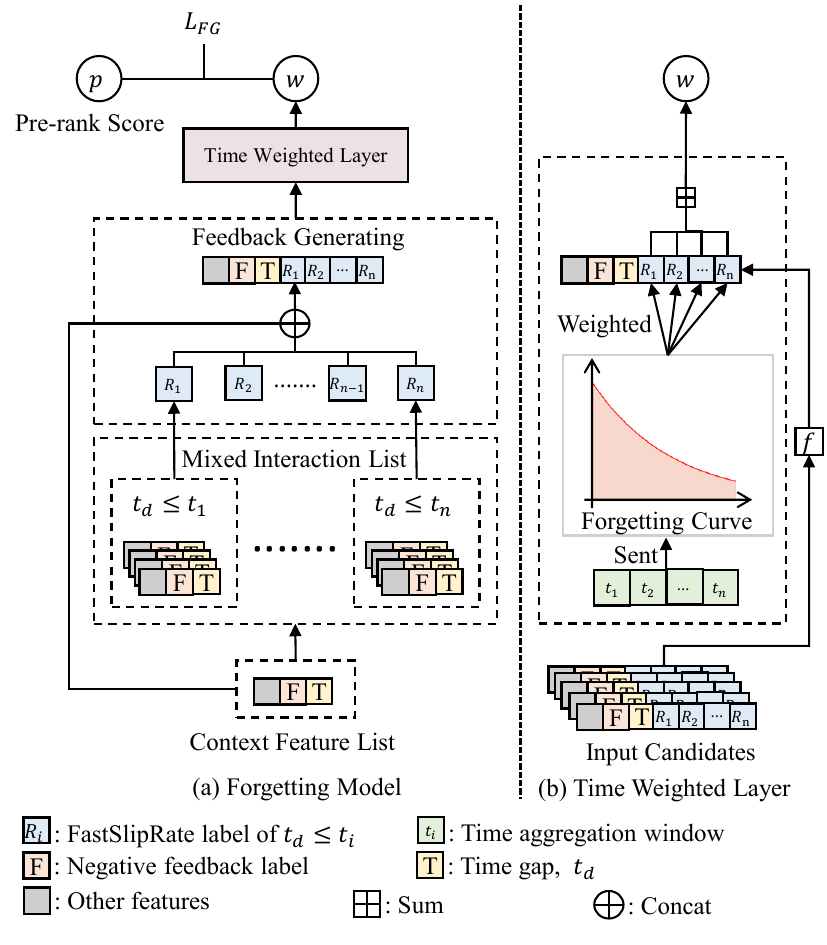}
  \caption{Overview Structure of Forgetting Model}
  \label{forgettingmodelimage}
  \Description{}
\end{figure}

The forgetting model is designed to optimize the \textit{recency} objective.
For short video platform, each exposure to the candidate at a specific time serves as a \textit{review} of the impression that the candidate leaves to the user, and each fast-slip action expresses users' explicit aversion towards the exposed candidate. The process of adopting the forgetting model is shown in Fig.~\ref{forgettingmodelimage}, consisting of three main parts: gathering interaction lists by time, feedback generating and a time weighted layer. 

To begin with, we set several time aggregation windows, $T=\{t_1,t_2,\cdots,t_n\}$, for calculating the \textit{FastSlipRate} in these time periods. For each interaction, the model separates the history interactions of the involved candidate into different lists according to the time gap ($t_d$) between history interactions and the current one, then counts the rate of negative feedback labels in each list by Eqn.~(\ref{R_i}) and outputs a list of  \textit{FastSlipRate}, $R=\{R_1,R_2,\cdots,R_n\}$. 
\vspace{-3pt}
\begin{equation}\label{R_i}
    R_j = \frac{N(t_d \leq t_j \ \text{and is negative})}{N(t_d \leq t_j)} .
\end{equation}

We should impose greater penalty on the candidate with greater memory strength and higher \textit{FastSlipRate}. To calculate the penalty weight, we design a time-weighted layer. The weight of $R_j$ is calculated by the Ebbinghaus Forgetting Curve with the corresponding time aggregation window $t_j$.
For the $i$-th candidate, we define its penalty weight:
\vspace{-3pt}
\begin{equation}\label{offlineweight}
    w_i = \sum^n_{j=1} \left( f(R_j) \cdot e^{-\frac{t_j}{S}} \right) ,
\end{equation}
where $T=\{t_1,t_2,\cdots,t_n\}$ is a series of time aggregation windows.
$f(R)$ is the pre-processing function for \textit{FastSlipRate}, mainly employed for dealing with abnormal values and optimizing value range distribution to enhance discrimination. 
For simplicity, we define $f(R)$ as an identity function in offline environment.  

Then we can calculate the loss of forgetting model by: 
\begin{equation}\label{penaltyloss}
    L_{FG} = -\sum_i w_i \cdot \log (1-p(\mathbf{x}_i)),
\end{equation}
where $p(\mathbf{x}_i)$ is the output of pre-ranking model.
It is noteworthy that we use $\log (1-p(\mathbf{x}_i))$ instead of $\log (p(\mathbf{x}_i))$ in Eqn.~(\ref{penaltyloss}), where we hope that more fast-slip penalty can be imposed on the candidates with higher ranking,
because in most cases these candidates will exert more influence to user experience.

\subsubsection{Pareto Solver}\label{Paretosolversection}

To find a solution $\theta$ satisfying KKT conditions mentioned in Definition \ref{Pareto stationarity}, we consider the following constrained minimization problem proposed by~\cite{MGDA2012}:
\begin{equation}\label{KKT}
\begin{aligned}
\mathop{\min}_{\alpha_1, \dots, \alpha_t} &\Bigg\Vert \sum_{i=1}^t \alpha_i \cdot \nabla_{\bm{\theta}} \mathcal{L} (\bm{\theta}) \Bigg\Vert^2_2, \\
s.t., &\sum^t_{i=1}\alpha_i = 1, \alpha_i \geq 0, for \ i = 1, \dots, t. 
\end{aligned}
\end{equation}

There are 2 situations for the solution to Problem (\ref{KKT}). 
If the solution to this optimization problem is 0, the resulting point satisfies the KKT conditions, which is Pareto stationary and also Pareto optimality under realistic and mild conditions~\cite{sener2018multi}. 
Otherwise, the solution offers a common descent direction to optimize $\bm{\theta}$ that benefits all the objectives~\cite{MGDA2012}. 
Proceeding along this line of thought, we propose Pareto optimization solver to optimize the solution to Pareto stationary, utilizing Frank-Wolfe algorithm~\cite{frankwolfe1956algorithm}. 
As shown in Alg. (\ref{ParetoSolver}), the preference weights $\bm{\alpha}=(\alpha_1, \dots, \alpha_t)$ can be generated to satisfy the KKT conditions aforementioned.

\begin{algorithm}
\caption{Pareto Optimization Solver~\cite{frankwolfe1956algorithm}}\label{ParetoSolver}
\SetKwRepeat{Do}{do}{while}
\SetKwInOut{Input}{input}\SetKwInOut{Output}{output}
\Input{$\nabla_{\bm{\theta}} \mathcal{L}_i(\bm{\theta})$, for $i\in \{1,\dots,t\}$

       }
\Output{weights for each loss function: $\bm{\alpha}=(\alpha_1, \dots, \alpha_t)$}
\textbf{Random Initialization}: $\bm{\alpha}=(\alpha_1, \dots, \alpha_t)$ satisfying the constraints in Problem (\ref{KKT})\\
\textbf{Precompute} $\bm{M}$:\\
\qquad $\bm{M}_{ij} = (\nabla_{\bm{\theta}} \mathcal{L}_i(\bm{\theta}))^T(\nabla_{\bm{\theta}} \mathcal{L}_j(\bm{\theta})),\quad \forall i,j \in \{1,\dots,t\}$;\\
\Repeat{$w^*$ converge or maximum iteration reaches}{
      $i^* = \arg\min_r \sum_i\alpha_i\bm{M}_{ri}$;\\
      $w^* = \arg\min_w (w\bm{e}_{i^*}+(1-w)\bm{\alpha})^T \bm{M} (w\bm{e}_{i^*}+(1-w)\bm{\alpha})$;\\
      \qquad \qquad \qquad \qquad \qquad \qquad$\uparrow$ Using Alg. (\ref{compute})\\ 
      $\bm{\alpha} = w^*\bm{e}_{i^*}+(1-w^*)\bm{\alpha}$; \qquad ($\bm{e}_{i^*}$ is the unit vector)
    }
\textbf{Return} $\bm{\alpha} = (\alpha_1, \dots, \alpha_t)$;
\end{algorithm}

\begin{algorithm}
\caption{Compute $w^*$}\label{compute}
\textbf{Solving} $\arg\min_{w\in [0,1]}\Vert wx_1+(1-w)x_2 \Vert^2_2$\\
\qquad $w^* = \frac{(x_2-x_1)^Tx_2}{\Vert x_1-x_2\Vert^2_2}$; \\
\qquad $w^* = \max(\min(w^*,1),0)$; \\
\textbf{Return} $w^*$;
\end{algorithm}

Therefore, during the training stage, we firstly generate losses for each objective by forward passing and then calculate their gradients which are prepared for the Pareto Solver. The output of the Solver is a series of scaling coefficients that will be weights for their specific objectives. After aggregating the total loss by weight-suming each objective loss, we calculate the new gradient of total loss and update our solution $\theta$. Finally, the updating loop ends with an expected solution $\theta$ that leads to Pareto stationary.

\subsection{Training Strategy}

In the training stage, the ranking model is fixed and offers relevant scores in the pre-ranking list. In details, for each impression, we sample 10 candidates and collect their rank scores given by the ranking model to train our pre-ranking model by $L_{LTR}$ in Sec.~\ref{ranking and pre-ranking}. In the testing stage, for each impression, we sample 100 candidates that are given to pre-ranking model to select the top 10 candidates to present. Then these candidates are sent to the ranking model to select the top 1 candidate as the final one clicked by users. During the whole process, the forgetting model calculates $L_{FG}$ that measures the effect of feedbacks which is mentioned in Sec.~\ref{FORGETTING MODEL}.

Finally, the total loss $L_{all}$ is proposed to be calculated in Eqn.~(\ref{Lall}). To better combine the two important objectives, a Pareto optimization solver mentioned in section \ref{Paretosolversection} is adopted to provide the preference weights, namely $\alpha_{LTR}$ and $\alpha_{FG}$. By minimizing $L_{all}$, PMORS optimizes all objectives in a Pareto efficient way.   
\begin{equation}\label{Lall}
    L_{all} = \alpha_{LTR}L_{LTR}+\alpha_{FG}L_{FG} .
\end{equation}

\section{Experiments}\label{experiment}

\subsection{Experiment Setup}
\subsubsection{Dataset}
KuaiRec dataset~\cite{gao2022kuairec} is a public dataset\footnote{\url{https://kuairec.com/}} with 12.5 million interactions between users and items which are collected from recommendation logs of Kuaishou, a video-sharing app. 
We adopt the definition of negative feedback and positive interaction from \cite{gao2022kuairec}.
The dataset is split into 90\% (for training), 5\% (for validating), 5\% (for testing) based on the timestamps of the samples. 

\subsubsection{Evaluation Metrics}
To fully estimate the overall performance of our model in the given dataset, we adopt NDCG@10,Recall@10\_1, CTR and FastSlipRate as metrics for PMORS. 

\textbf{NDCG@10. } NDCG@10 measures the quality of ranking results by considering both the relevance and the position of the items in the top 10 positions. 

\textbf{Recall@10\_1. }Recall@10\_1 measures the average recall ratio of top-1 candidate recommended by the ranking stage in the top-10 candidates suggested by the pre-ranking stage, evaluating the ranking \textit{consistency}. Recall@10\_1 prioritizes the ultimate individual outcome of the ranking stage, rendering it more significant than NDCG@10 in industrial applications. It is shown in Eqn.~(\ref{recall}) and the $N$ represents the number of recommendation requests.
\vspace{-3pt}
\begin{equation}\label{recall}
    \text{Recall}@10\_1 = \frac{1}{N}\sum\limits_{i=1}^{N}{\big|\text{Top}10_{\text{pre-rank}}\cap \text{Top}1_{\text{rank}}\big|} .
\end{equation}
\vspace{-3pt}

\textbf{CTR. }CTR (Click-Through Rate) is a metric used to measure the effectiveness of candidate clicks and is commonly employed to evaluate the performance of recommendation systems. 

\textbf{FastSlipRate. }We develop the FastSlipRate metric in order to show the possibility of fast-slip action on the impression candidates by users, indicating the reduction of negative feedbacks on short video platform. Similar to CTR, the formula is shown as follows:
\vspace{-3pt}
\begin{equation}\label{FastSlip}
    \text{FastSlipRate} = \frac{{\text{Number of FastSlips}}}{{\text{Number of Impressions}}} .
\end{equation}
When it comes to our online deployment in Sec.~\ref{Online result}, we also utilize the Gross-Merchandise-Volume (GMV) since it serves as a direct view of profit in recommender system.

\subsubsection{Baselines}\label{baselines}
We compare our method with baselines from three domains: (a) typical Learning to Rank methods (RankNet, LambdaMART) (b) multi-task methods (ESMM, MMOE) and (c) Pareto approaches (PO-EA, PE-LTR). 
\begin{itemize}
    \item \textbf{RankNet} ~\cite{2005RankNet} is a supervised machine-learning method using gradient descent and probabilistic ranking cost function to model each candidate pair with neural network. 
    \item \textbf{LambdaMART} ~\cite{2010LambdaMART}  is a popular listwise LTR model which linearly combines the boosted regression trees to construct the overall ranking function and optimizes model parameters through gradient descent.
    \item  \textbf{ESMM} ~\cite{2018ESMM} is a multi-task model for CTR and CVR predictions, which focuses on the various sample spaces towards each task. In our experiment, we set \textit{consistency} and \textit{recency} as the two tasks for ESMM model.
    \item  \textbf{MMOE} ~\cite{2018MMOE} focuses on modelling trade-offs between inter-task relationships and task-specific objectives which are represented by the shared expert submodels across all tasks and the task-specific towers.
    \item \textbf{PO-EA}~\cite{2014POEA} is a Pareto-efficient multi-objective recommendation method. It combines different elementary algorithms, resulting in a hybrid one. The combination weights are generated with evolutionary algorithm. 
    \item \textbf{PE-LTR} ~\cite{2019Pareto} is a general framework for generating Pareto efficient recommendations. It proposes a two-step Pareto efficient optimization algorithm to learn the preference weights for coordinating differential formula of objectives like CTR and GMV in recommender system.
    \item \textbf{Contrastive Learning (CL)} is another negative feedback penalty method with indirect penalty loss inspired by~\cite{zhou2020s3}. Given a candidate with disturbed negative feedback rate, the loss guides model to adjust the score accordingly. It adopts the same Pareto solver and forgetting model with PMORS.
\end{itemize}

\subsubsection{Hyperparameter settings}
For fair comparison in the following experiments, the LTR baselines and pre-ranking part in PMORS share the exact same network structure and hyperparameters. We set the embedding size as 16 and the hidden parameters in three-layer neural network are decided as $\left[128,64,32 \right]$. We form $T$ as \{10min,3h,1day,7days\} in the forgetting model, counting the historical fast-slip actions to calculate the $R_j$ in Eqn.~(\ref{offlineweight}) on each time aggregation window. We select the Adam\cite{kingma2014adam} as the general training optimizer which deals with losses multiplied by preference weights given by the Pareto optimization solver, and set the learning rate to 0.001.

\subsection{Offline Results}
We calculate NDCG@10,FastslipRate, Recall@10\_1 and CTR metrics on KuaiRec dataset to examine the overall performance. We also conduct ablation study by fixing the preference weights.  Furthermore, we verify the wide scalability that PMORS has when being equipped with different ranking models. Finally, we explore the influence of relative memory strength $S$ in the forgetting model.
\begin{table*}
  \caption{Overall Performance Results on KuaiRec Dataset. Impr. indicates the relative improvement compared to RankNet. All results are statistically significant with $p < 0.05$.}
  \label{offline}
\begin{tabular}{@{}c|cccccccc@{}}
\toprule
Model      & NDCG@10         & Impr.           & FastSlipRate      & Impr.             & CTR               & Impr.            & Recall@10\_1      & Impr.            \\ \midrule
RankNet    & {\ul 0.6974}    & -               & 0.1458\%          & -                 & 0.3458\%          & -                & 1.7293\%          & -                \\
LambdaMART & \textbf{0.7067} & \textbf{1.32\%} & 0.1662\%          & 14.01\%           & 0.3686\%          & 6.59\%           & 1.9440\%          & 12.42\%          \\ \midrule
ESMM       & 0.5107          & -26.77\%        & \textbf{0.0218\%} & \textbf{-85.01\%} & 0.2932\%          & -15.22\%         & 1.2962\%          & -25.05\%         \\
MMOE       & 0.5515          & -20.92\%        & {\ul 0.0710\%}    & {\ul -51.33\%}    & 0.2987\%          & -13.63\%         & 1.4986\%          & -13.34\%         \\ \midrule
PO-EA      & 0.6131          & -12.09\%        & 0.1024\%          & -29.78\%          & 0.3804\%          & 9.99\%           & 2.1470\%          & 24.16\%          \\
PE-LTR     & 0.6507          & -6.70\%         & 0.1021\%          & -29.95\%          & {\ul 0.3831\%}    & {\ul 10.77\%}    & {\ul 2.2530\%}    & {\ul 30.28\%}    \\ \midrule
CL         & 0.6907          & -0.96\%         & 0.1554\%          & 6.61\%            & 0.3582\%          & 3.57\%           & 2.0632\%          & 19.31\%          \\
PMORS    & 0.6650          & -4.65\%         & 0.1064\%          & -26.99\%          & \textbf{0.3851\%} & \textbf{11.35\%} & \textbf{2.3386\%} & \textbf{35.24\%} \\ \bottomrule
\end{tabular}
\end{table*}

\subsubsection{Overall Performance}
Tab.~\ref{offline} compares PMORS and baselines in terms of \textit{consistency} metric (NDCG@10), \textit{recency} metric (FastSlipRate) and system performance (CTR,Recall@10\_1). From Tab.~\ref{offline}, we draw several conclusions.  

Although PMORS does not present the greatest improvement on any single objective, it achieves the best system performance among all baselines.  We attribute the improvement to two aspects: the effectiveness of \textit{recency} objective and Pareto optimization for combining two objectives. 

Non-Pareto optimization models focus on a single objective to which they bring remarkable improvement, but they fail on another objective and present weak system performance. LTR models are so strongly intended to optimize the \textit{consistency} objective that they suffer high FastSlipRate and their system performance is low compared to Pareto-based models. For multi-task models, they have paid huge effort to reduce the FastSlipRate for optimizing \textit{recency} objective. However, the discouraging performance both in \textit{consistency} objective and system performance indicates the seesaw phenomenon in recommender system. Although MMOE can relieve that phenomenon by good expert gates for evaluating inter-task relationships, its performance is significantly frustrating compared to Pareto-based models.  

Pareto-based optimization models can bring Pareto optimal solutions to better weight the two objectives. After setting its objectives as \textit{consistency} and \textit{recency}, PE-LTR model can present comparable system performance despite the difference in generating Pareto solutions based on scalarization method. The low performance in contrastive learning model indicates the disadvantage of indirect penalty. PO-EA model does not present expected results because its evolutionary algorithm is less capable to generate Pareto-efficient solutions, which is similar to the conclusion in ~\cite{2019Pareto}.

\subsubsection{Ablation Study}
To enhance the effectiveness of Pareto optimization and forgetting model, we conducted an ablation study where we substituted the Pareto optimization solver in PMORS with a range of fixed $\alpha_{LTR}$, specifically $\left\{ 0.5,0.6,0.7,0.8,0.9,1.0 \right\}$. If the fixed $\alpha_{LTR}$ is 0.6, the total loss will be $L_{all} = 0.6L_{LTR}+0.4L_{FG}$. 
For $\alpha_{LTR}$ less than 0.5, the model will collapse due to  float underflow since the $L_{FG}$ (loss of a simpler task than LTR) approaches 0. Note that the forgetting model is blocked when $\alpha_{LTR}=1$.
NDCG@10 and FastSlipRate of fixed weights' samples are shown in Fig.~\ref{ablation} as the blue points and our model (red) is also presented for comparison. Tab.~\ref{AblationStudy} shows the metrics of system performance

Certainly, as the weight we deliver to the \textit{consistency} objective increases, NDCG@10 improves while FastSlipRate becomes worse. Compared to models with fixed weights, PMORS finds a better trade-off between \textit{consistency} and \textit{recency}, showing the optimality of Pareto algorithm.
The introduction of forgetting model improves the system performance as the models with $\alpha_{LTR} \in \{0.8, 0.9\}$ outperform the one with forgetting model blocked. However, excessive weights on $L_{FG}$ in turn harms the whole system.
PMORS reaches the highest system performance on both metrics, indicating the effectiveness of Pareto solver.

\begin{table}[]
    \caption{System Performance in Ablation Study}
    \label{AblationStudy}
\begin{tabular}{@{}ccccc@{}}
\toprule
$\alpha_{LTR}$ & CTR      & Impr.      & Recall@10\_1 & Impr.      \\ \midrule
1  & 0.3458\% & -          & 1.7293\%     & -          \\ \midrule
0.9    & 0.3686\% & 6.5897\%   & 2.1135\%     & 22.2166\%  \\
0.8    & 0.3557\% & 2.8388\%   & 1.8627\%     & 7.7124\%   \\
0.7    & 0.3380\% & -2.2590\%  & 1.6402\%     & -5.1512\%  \\
0.6    & 0.3079\% & -10.9656\% & 1.4833\%     & -14.2229\% \\
0.5    & 0.2987\% & -13.6297\% & 1.2572\%     & -27.2976\% \\ \midrule
PMORS  & 0.3851\% & 11.3451\%  & 2.3386\%     & 35.2355\%  \\ \bottomrule
\end{tabular}
\end{table}

\subsubsection{Scalability.}
\begin{figure*}
    \centering 
    \setlength{\abovecaptionskip}{0pt}
    \subfigbottomskip=2pt
    \subfigcapskip=0pt
	\subfigure[NDCG@10]{
		\includegraphics[width=0.24\linewidth]{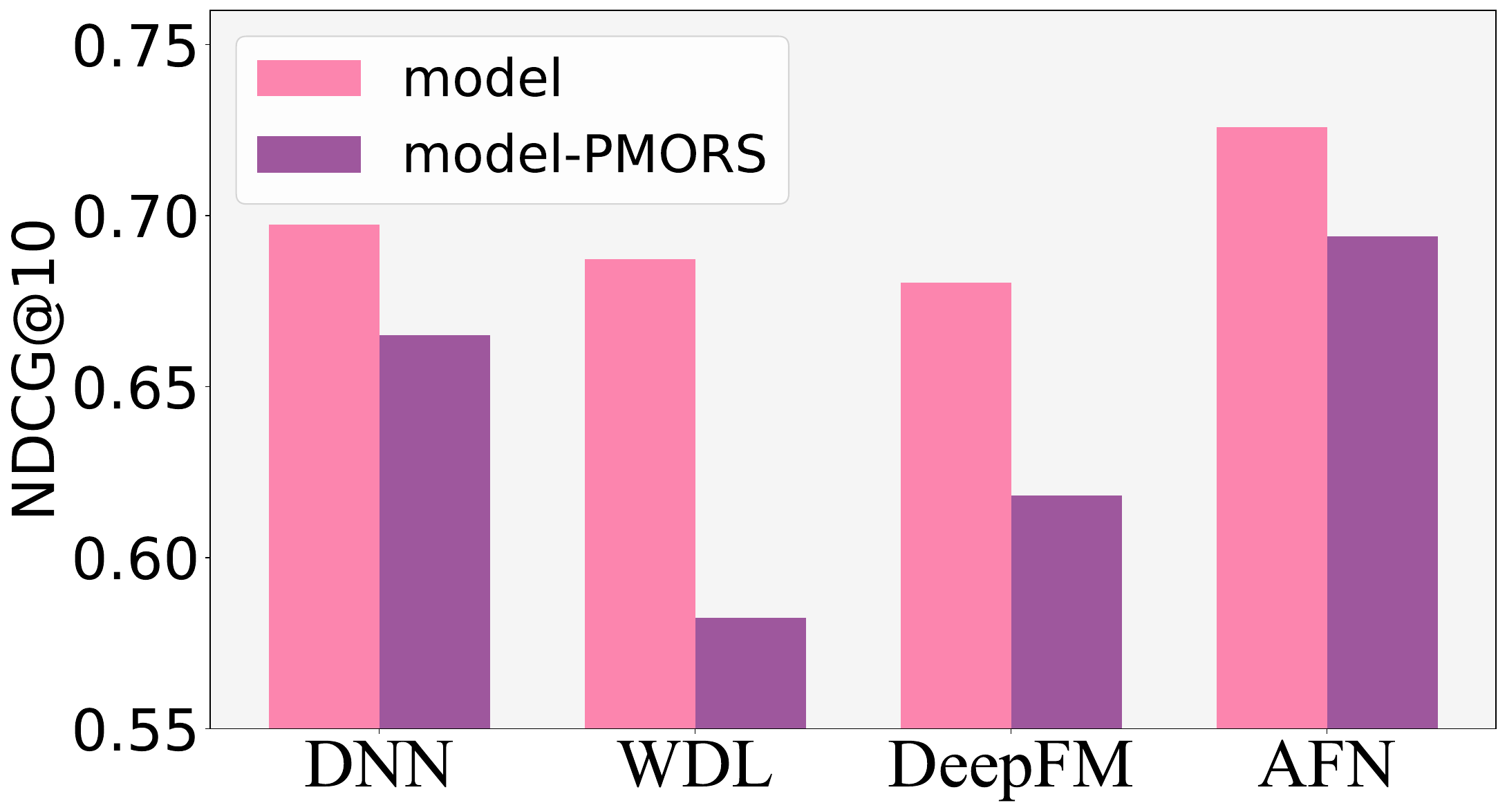}}
	\subfigure[FastSlipRate]{
		\includegraphics[width=0.24\linewidth]{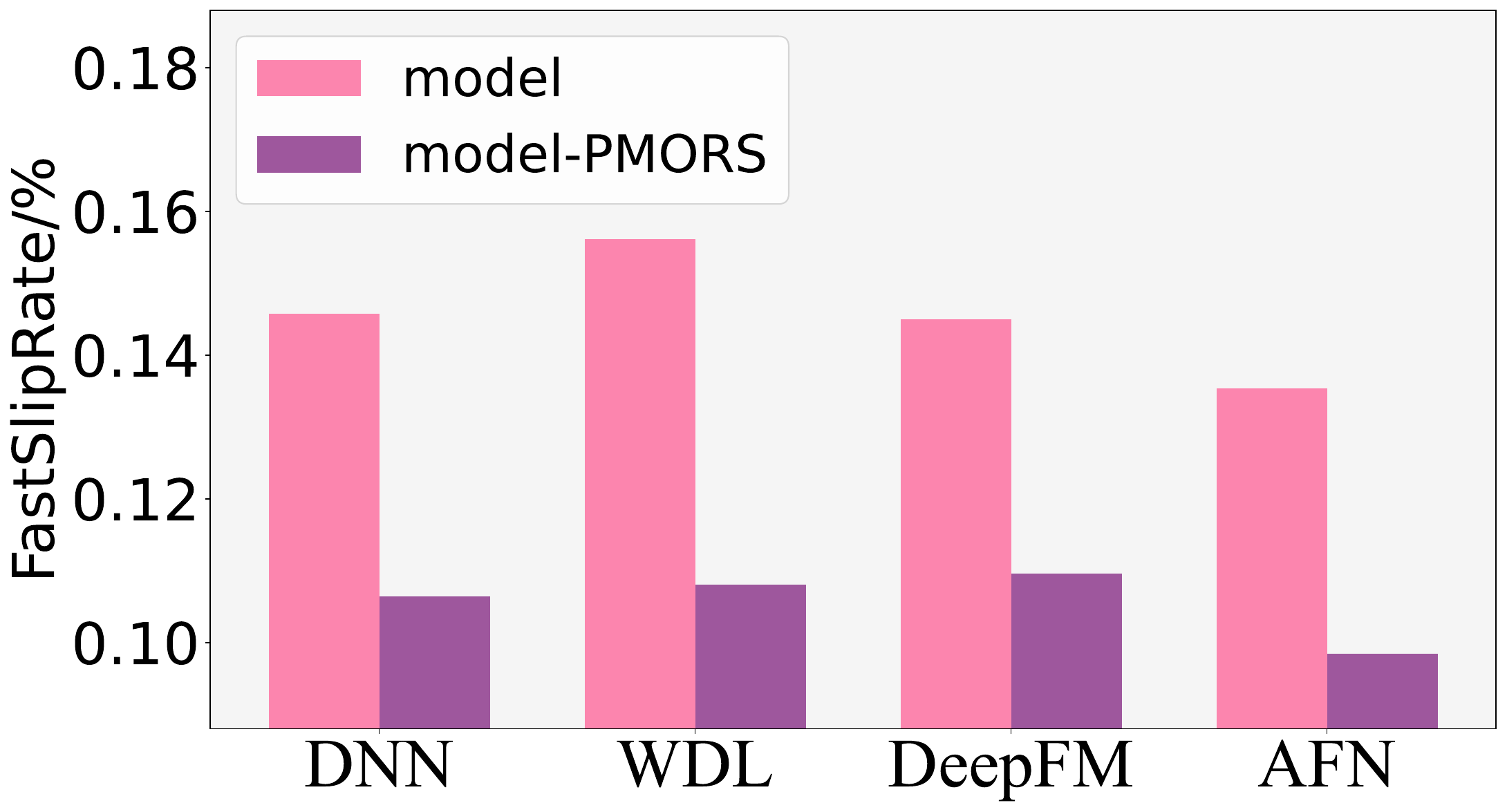}}
	\subfigure[CTR]{
		\includegraphics[width=0.24\linewidth]{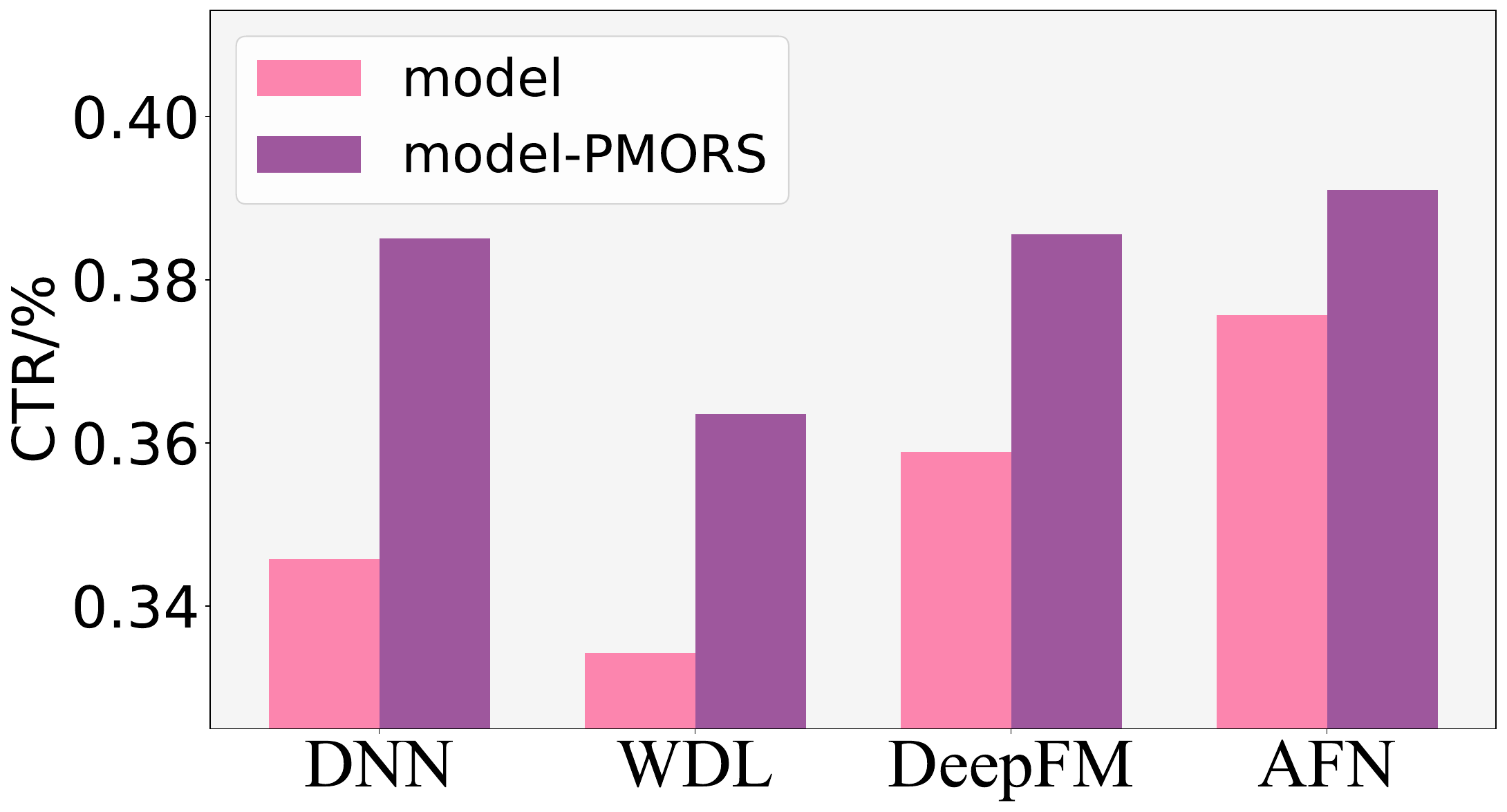}}
	\subfigure[Recall@10\_1]{
		\includegraphics[width=0.24\linewidth]{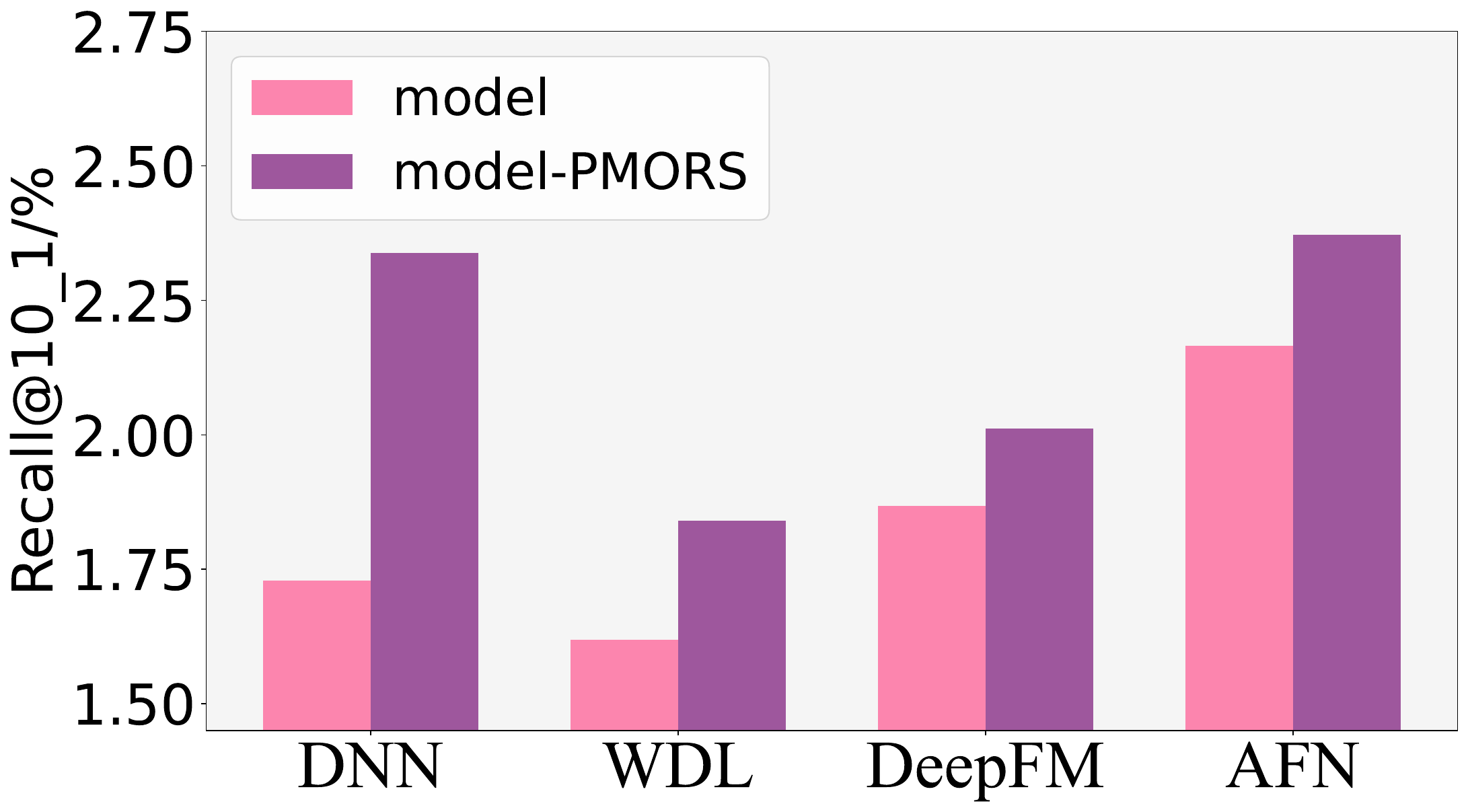}}
	\caption{Scalability Results}
 \label{Scalability}
\end{figure*}

PMORS does not rely on specific structures of pre-ranking models and can work as long as the pre-ranking model has gradients. 
We conduct experiments to show the scalability of PMORS in terms of model selection.
We use Deep Neural Network (DNN), Wide\&Deep (WDL)~\cite{widedeep}, DeepFM~\cite{deepfm} and Adaptive Factorization Network (AFN)~\cite{afn} as the pre-ranking model in PMORS and access the results.
DNN model is the RankNet mentioned in Sec.~\ref{baselines} and has a same structure with the deep component in the Wide\&Deep model. 
DeepFM is a neural network that combines the power of factorization machines and deep learning.
AFN learns arbitrary-order cross features adaptively from data. 
The results can be seen in Fig.~\ref{Scalability}.

The use of PMORS on each model leads to a significant improvement in system performance compared to the original one, while causing an acceptable decrease in NDCG@10.
Among these PMORS models, AFN-PMORS performs better than the other three models, and DeepFM-PMORS comes in second.
The results are explainable because AFN adaptively learns arbitrary-order cross features, which guarantees better performance than the others.
Similarly, DeepFM leverages the power of factorization machines combined with neural networks, enabling it to capture more relationships in features compared to Wide\&Deep model.
In conclusion, PMORS can adapt itself to different pre-ranking models with excellent scalability, and stronger models can achieve better results accordingly.

\subsubsection{Experiment about Relative Memory Strength}
As mentioned in Sec.~\ref{Ebbinghauscurves}, the relative memory strength is an unique value decided by the characteristics of original content to be memorized. To make it clear, we define $L$ as the memory rate after one day, $L = R(1) = e^{\frac{1}{-S}}$.  Instead of changing $S$ directly, we adjust $L$ in a list ($L = \left[ 0.1,0.2,\cdots,0.9 \right]$) to test the performance of our PMORS.

\begin{figure*}[h]
        \centering        
	\begin{minipage}[t]{0.32\textwidth}
 \vspace{0pt}
        \includegraphics[width=\textwidth]{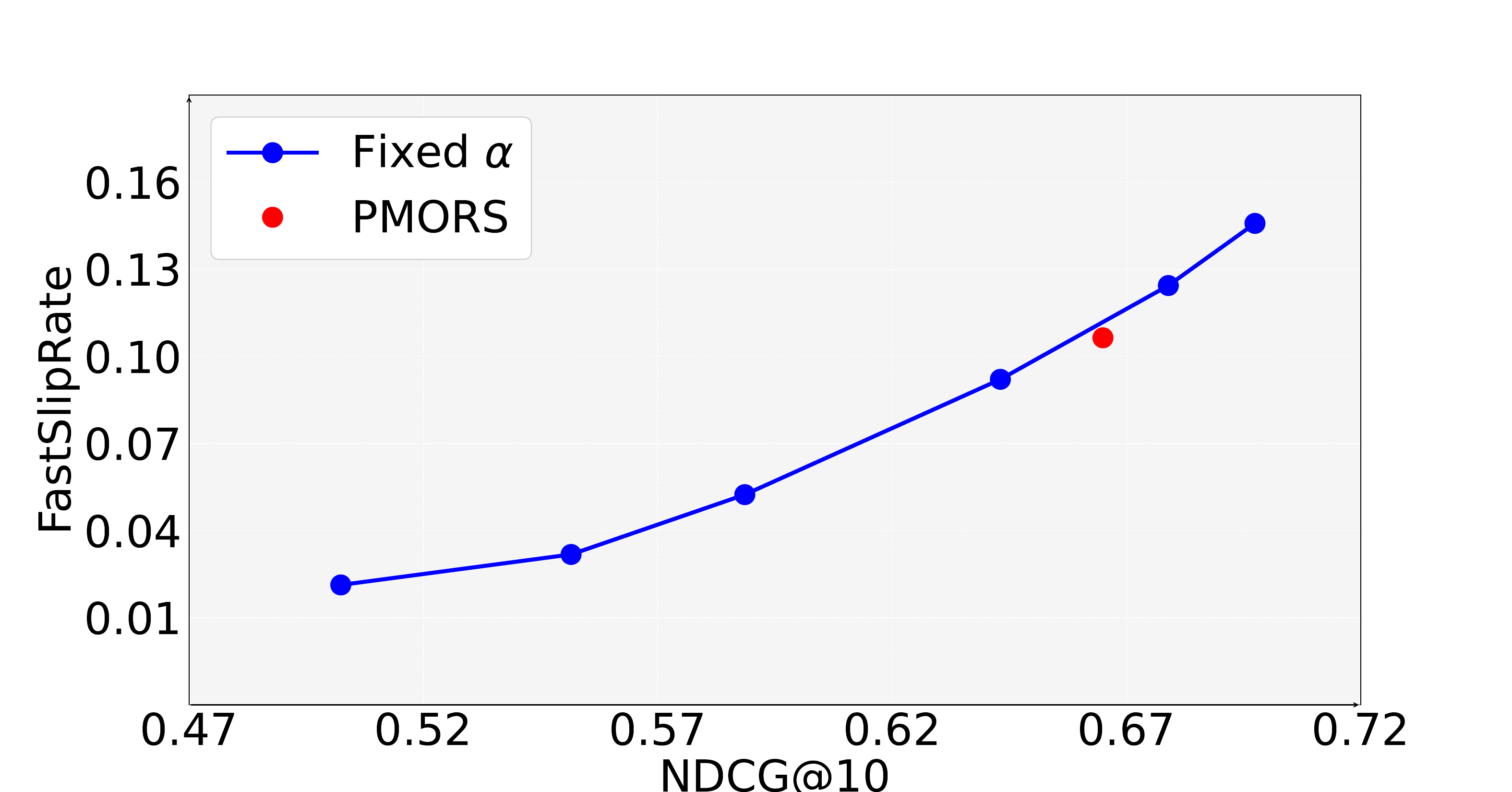}
        \caption{Consistency and Recency in Ablation Study}
        \label{ablation}
	\end{minipage}
        \hspace{-0.5cm}
        \begin{minipage}[t]{0.64\textwidth}
         \vspace{-5pt}
	\subfigure[NDCG,FastSlipRate]{
		\includegraphics[width=0.48\textwidth]{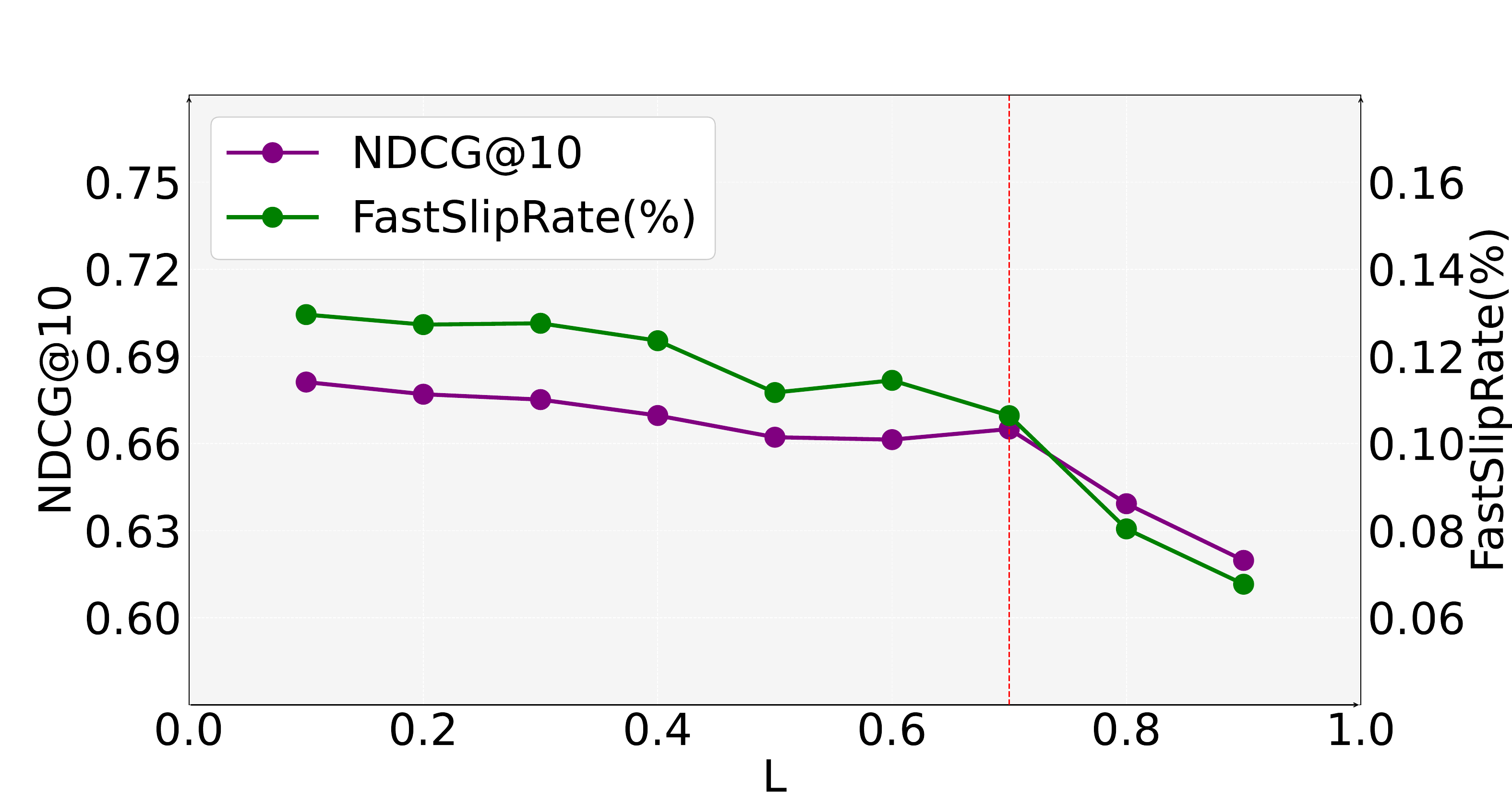}
        }
	\subfigure[CTR,Recall@10\_1]{
		\includegraphics[width=0.48\textwidth]{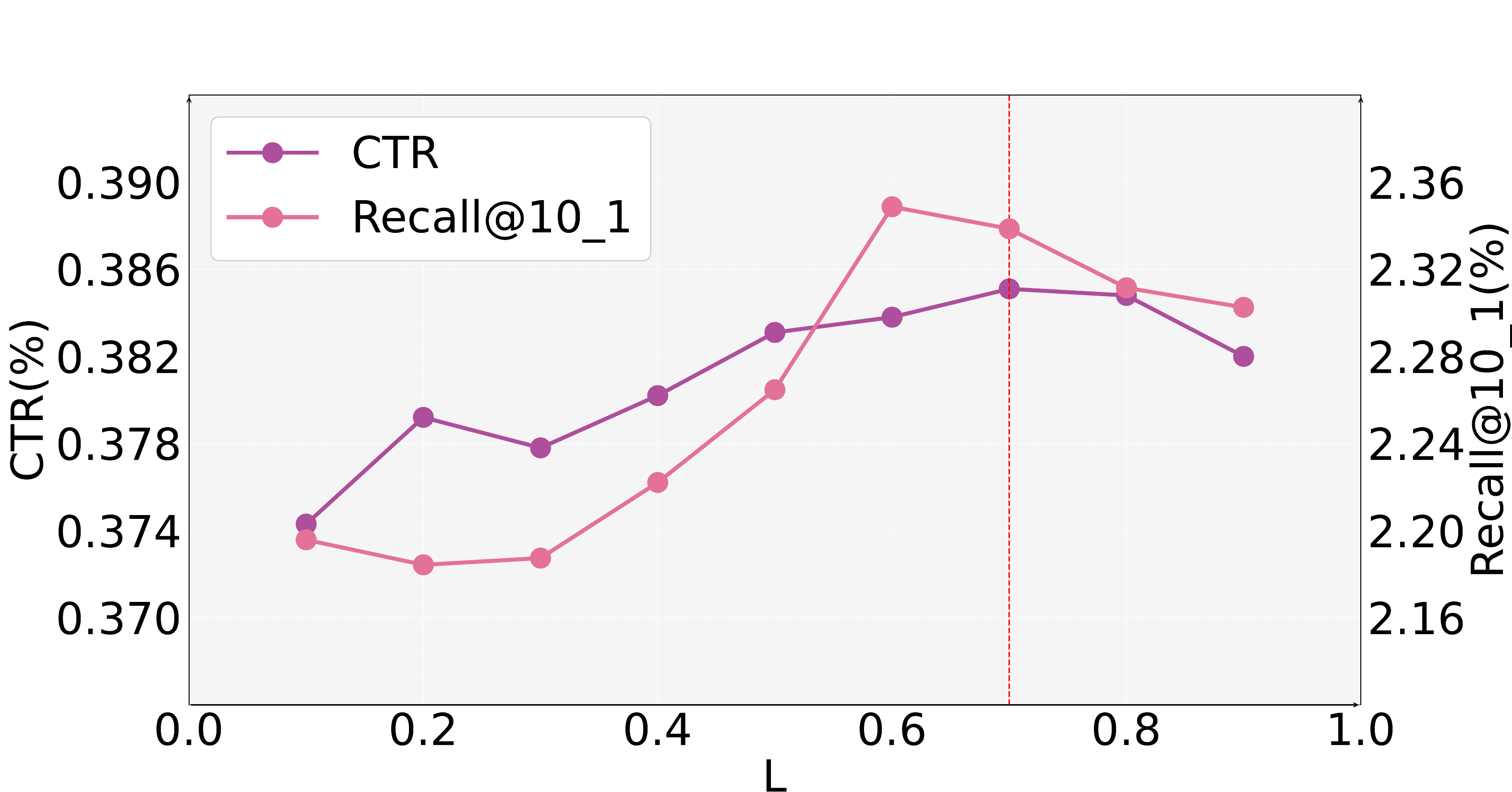}}
  \vspace{-14pt}
        \caption{Evaluation Metrics with Different $L$}
        \label{MemoryStrength}
        
        \end{minipage}
\end{figure*}

In Fig.~\ref{MemoryStrength}, it can be observed that both NDCG@10 and FastSlipRate are in decline with the increase of L, which is in line with the counter-balance of \textit{consistency} and \textit{recency}.
The metrics of system performance are suboptimal when $L$ is either too low or too high. Lower $L$ implies forgetful experiences to negative feedback while higher $L$ means unforgettable responses. The figure shows that there exists a most suitable $L$ between 0 and 1, corresponding to the relative memory strength of this type of content. 

After comprehensive evaluation based on these metrics, we set $L = 0.7$ that is produced by $S=6.46$ in our offline experiment. 
For our online deployment on WeChat Channels, we perform a more fine-grained categorization of candidates and allocated them different values of $S$.

\subsection{Online Deployment}\label{Online result}

Intuitively, users demonstrate different levels of relative memory strength for different types of information. Therefore, we utilize unique online information to allow for a more fine-grained segmentation of fast-slip action.
Specifically, to perform better in online environment, we divide fast-slip actions into 3 dimensions, i.e., material fingerprint, item fingerprint, and secondary industry. 
Material fingerprint is a cluster ID generated by information extraction models and clustering models based on the material information of the candidate, mainly about videos, figures and contexts. 
Similarly, item fingerprint is another cluster ID based on the item information of the candidate, mainly about categories, item IDs and SPUs (Standard Product Unit). 
Secondary industry is the subdivision of industries within the Tencent advertising ecosystem.

Correspondingly, each dimension has a relative memory strength for users, denoted by $S_m$ for the $m$-th dimension.
For the $i$-th candidate, we adjust its penalty weight to
\begin{equation}\label{onlineweight}
    w_i = \sum^M_m \lambda_m \sum^n_{j=1} \left( f(R_j) \cdot e^{-\frac{t_j}{S_m}} \right),
\end{equation}
where $\lambda$ is the weight of these dimensions, $\lambda_m \in (0,1),\  \sum_m \lambda_m = 1$. $f(R)$ is adjusted precisely to enhance the discrimination.

To evaluate the performance of our model in real-world production environments, the online A/B test is conducted on the WeChat Channels platform in May, 2023 and the results are shown in Tab.~\ref{online}, where the baseline is our last online LTR model. 
Our model performs better on all metrics, especially on consistency (Recall@10\_1).
Now our model has been successfully deployed online to serve the main traffic of WeChat Channels and and demonstrated promising results.

\begin{table}
  \caption{Online Deployment Results.}
  \label{online}
  \begin{tabular}{ccccc}
    \toprule
             GMV     & FastSlipRate & FastSlipAction & CTR     & Recall@10\_1 \\
    \midrule
 
             +1.45\% &   -0.63\%    &   -0.50\%      & +0.89\% &   +3.0\%    \\
    \bottomrule
  \end{tabular}
\end{table}

\section{Discussion}
In the discussion part, we are going to share an interesting topic about the weight gathering method of the forgetting model. Take fastslip as an example. Since our aim is to measure the effect of this action, our model absorbs all interactions with the specific candidate in the given time aggregation windows and the weight $w$ will be applied equally for all users. Different from our generalized method, we also consider whether a more personalized method of weight gathering can be applied. 

\begin{figure}[htbp]
  \centering
  \includegraphics[width=1.05\linewidth]{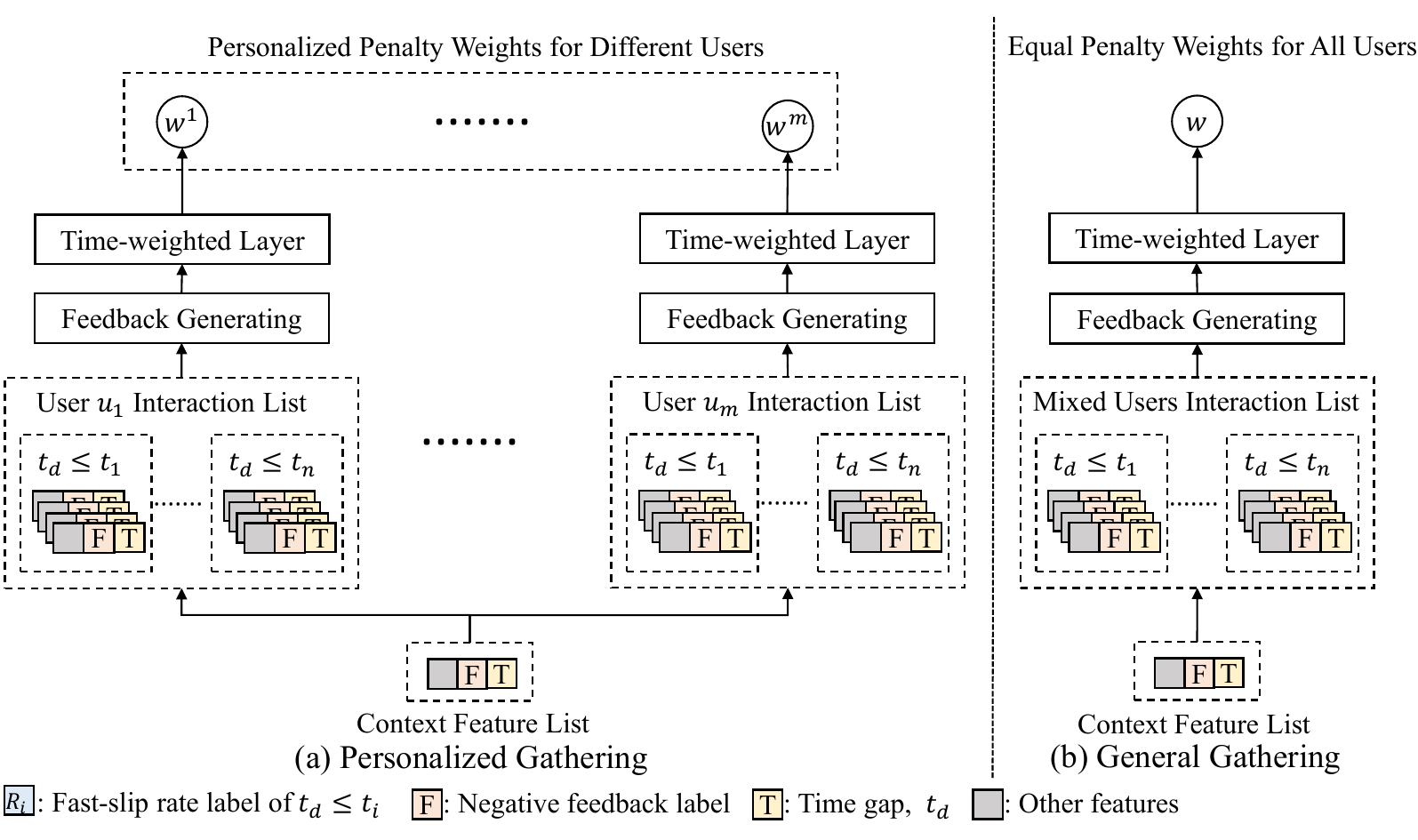}
  \caption{Personalized Gathering and General Gathering}
  \label{Personal general}
  \Description{}
\end{figure}

It focuses on each user's interacting history which will be separated timely for calculating the $R_i$ in the given time aggregation window $t_i$. Then $w^m$ will be unique penalty weight to user $m$ because it considers his own interaction list, so will the final $L_{FG}$. The structures of two weight aggregating methods are shown in Fig.~\ref{Personal general}.

Comparing the two methods, it is clear that the personalized gathering provides statistics varied by users for each candidate, which are probably to bring more accurate $w$ for user feedback modelling. However, when applying the two methods in the industry scenario, another serious issue arises that the personalized gathering suffers from the shortage of personal interactions with the same candidate as shown in Tab.~\ref{shortage}. We count the amount of samples that are included in the pre-ranking task (divided into different granularity of fast-slip action, including item fingerprint and material fingerprint) and calculate their coverage rates in the given time aggregation windows ($T=\{ \text{10min,3h,1day,7days} \}$). 
\begin{table}[htbp]
  \caption{Negative Feedback Feature Coverage Rate (\%)}
  \label{shortage}
  \begin{tabular}{cccccc}
    \toprule
             Granularity & Method  & 10min & 3h & 1day     & 7days \\
    \midrule
 
             Item Fingerprint & Personalized &   0.04    &   0.30     & 1.00 &   1.80    \\
             Item Fingerprint  & General & 89 & 89 & 89 & 89\\
             Material Fingerprint & Personalized & 0.0001 & 0.02 & 0.08 & 0.15 \\
             Material Fingerprint & General & 91 & 95 & 95 & 95 \\
    \bottomrule
  \end{tabular}
\end{table}

Result shows that the personalized gathering has little coverage rate on all windows. With low coverage rate, it is hard for the model to optimize $L_{FG}$ since most of FastSlipRates are default values that we set to 0, useless for the willingly personalized forgetting model. It is not surprising because the recommender system is equipped with a good rank model to produce reasonable impression candidate list to users, which means the candidate that was fast-slipped by the user in the past can be significantly hard to be shown to the same user again. 

\section{Conclusion}

This paper introduces PMORS, a proposed model for muti-objective recommender systems with negative feedback.
We investigate methods for effectively handling negative feedback in recommender systems by incorporating the forgetting curve to apply suitable penalties.
Additionally, we utilize Pareto theory to ensure the reduction of negative feedback in the multi-objective recommender system while maintaining its utility.
The experimental results substantiate the effectiveness and scalability of the proposed model.
Furthermore, the online deployment of PMORS on WeChat Channels successfully reduces negative feedback without compromising recommendation performance, resulting in favorable economic benefits with a lift of up to +1.45\% GMV.

\bibliographystyle{ACM-Reference-Format}
\bibliography{sample-base}


\begin{thebibliography}{75}


\ifx \showCODEN    \undefined \def \showCODEN     #1{\unskip}     \fi
\ifx \showDOI      \undefined \def \showDOI       #1{#1}\fi
\ifx \showISBNx    \undefined \def \showISBNx     #1{\unskip}     \fi
\ifx \showISBNxiii \undefined \def \showISBNxiii  #1{\unskip}     \fi
\ifx \showISSN     \undefined \def \showISSN      #1{\unskip}     \fi
\ifx \showLCCN     \undefined \def \showLCCN      #1{\unskip}     \fi
\ifx \shownote     \undefined \def \shownote      #1{#1}          \fi
\ifx \showarticletitle \undefined \def \showarticletitle #1{#1}   \fi
\ifx \showURL      \undefined \def \showURL       {\relax}        \fi
\providecommand\bibfield[2]{#2}
\providecommand\bibinfo[2]{#2}
\providecommand\natexlab[1]{#1}
\providecommand\showeprint[2][]{arXiv:#2}

\bibitem[Aghaabbasi et~al\mbox{.}(2023)]%
        {2023bayesianhyperparameter}
\bibfield{author}{\bibinfo{person}{Mahdi Aghaabbasi}, \bibinfo{person}{Mujahid Ali}, \bibinfo{person}{Micha{\l} Jasi{\'n}ski}, \bibinfo{person}{Zbigniew Leonowicz}, {and} \bibinfo{person}{Tom{\'a}{\v{s}} Nov{\'a}k}.} \bibinfo{year}{2023}\natexlab{}.
\newblock \showarticletitle{On Hyperparameter Optimization of Machine Learning Methods Using a Bayesian Optimization Algorithm to Predict Work Travel Mode Choice}.
\newblock \bibinfo{journal}{\emph{IEEE Access}}  \bibinfo{volume}{11} (\bibinfo{year}{2023}), \bibinfo{pages}{19762--19774}.
\newblock


\bibitem[Belete and Huchaiah(2022)]%
        {gridsearch}
\bibfield{author}{\bibinfo{person}{Daniel~M. Belete} {and} \bibinfo{person}{Manjaiah~D. Huchaiah}.} \bibinfo{year}{2022}\natexlab{}.
\newblock \showarticletitle{Grid Search in Hyperparameter Optimization of Machine Learning Models for Prediction of HIV/AIDS Test Results}.
\newblock \bibinfo{journal}{\emph{International Journal of Computers and Applications (IJCA)}} \bibinfo{volume}{44}, \bibinfo{number}{9} (\bibinfo{year}{2022}), \bibinfo{pages}{875--886}.
\newblock


\bibitem[Bischl et~al\mbox{.}(2023)]%
        {2023hyperparameter}
\bibfield{author}{\bibinfo{person}{Bernd Bischl}, \bibinfo{person}{Martin Binder}, \bibinfo{person}{Michel Lang}, \bibinfo{person}{Tobias Pielok}, \bibinfo{person}{Jakob Richter}, \bibinfo{person}{Stefan Coors}, \bibinfo{person}{Janek Thomas}, \bibinfo{person}{Theresa Ullmann}, \bibinfo{person}{Marc Becker}, \bibinfo{person}{Anne-Laure Boulesteix}, {et~al\mbox{.}}} \bibinfo{year}{2023}\natexlab{}.
\newblock \showarticletitle{Hyperparameter Optimization: Foundations, Algorithms, Best Practices, and Open Challenges}.
\newblock \bibinfo{journal}{\emph{Wiley Interdisciplinary Reviews: Data Mining and Knowledge Discovery (WIREs)}} \bibinfo{volume}{13}, \bibinfo{number}{2} (\bibinfo{year}{2023}), \bibinfo{pages}{e1484}.
\newblock


\bibitem[Burges et~al\mbox{.}(2005)]%
        {2005RankNet}
\bibfield{author}{\bibinfo{person}{Chris Burges}, \bibinfo{person}{Tal Shaked}, \bibinfo{person}{Erin Renshaw}, \bibinfo{person}{Ari Lazier}, \bibinfo{person}{Matt Deeds}, \bibinfo{person}{Nicole Hamilton}, {and} \bibinfo{person}{Greg Hullender}.} \bibinfo{year}{2005}\natexlab{}.
\newblock \showarticletitle{Learning to Rank Using Gradient Descent}. In \bibinfo{booktitle}{\emph{International Conference on Machine Learning(ICML)}}. \bibinfo{pages}{89--96}.
\newblock


\bibitem[Candelieri et~al\mbox{.}(2023)]%
        {2023distributional}
\bibfield{author}{\bibinfo{person}{Antonio Candelieri}, \bibinfo{person}{Andrea Ponti}, \bibinfo{person}{Ilaria Giordani}, \bibinfo{person}{Anna Bosio}, {and} \bibinfo{person}{Francesco Archetti}.} \bibinfo{year}{2023}\natexlab{}.
\newblock \showarticletitle{Distributional Learning in Multi-Objective Optimization of Recommender Systems}.
\newblock \bibinfo{journal}{\emph{Journal of Ambient Intelligence and Humanized Computing (JAIHC)}} \bibinfo{volume}{14}, \bibinfo{number}{8} (\bibinfo{year}{2023}), \bibinfo{pages}{10849--10865}.
\newblock


\bibitem[Chen et~al\mbox{.}(2023)]%
        {2023revisiting}
\bibfield{author}{\bibinfo{person}{Chong Chen}, \bibinfo{person}{Weizhi Ma}, \bibinfo{person}{Min Zhang}, \bibinfo{person}{Chenyang Wang}, \bibinfo{person}{Yiqun Liu}, {and} \bibinfo{person}{Shaoping Ma}.} \bibinfo{year}{2023}\natexlab{}.
\newblock \showarticletitle{Revisiting Negative Sampling vs. Non-sampling in Implicit Recommendation}.
\newblock \bibinfo{journal}{\emph{ACM Transactions on Information Systems(TOIS)}} \bibinfo{volume}{41}, \bibinfo{number}{1} (\bibinfo{year}{2023}), \bibinfo{pages}{1--25}.
\newblock


\bibitem[Chen et~al\mbox{.}(2021)]%
        {2021fuzzy}
\bibfield{author}{\bibinfo{person}{Jianrui Chen}, \bibinfo{person}{Yanqing Lu}, \bibinfo{person}{Fanhua Shang}, {and} \bibinfo{person}{Yuyang Wang}.} \bibinfo{year}{2021}\natexlab{}.
\newblock \showarticletitle{A Fuzzy Matrix Factor Recommendation Method with Forgetting Function and User Features}.
\newblock \bibinfo{journal}{\emph{Applied Soft Computing (ASC)}}  \bibinfo{volume}{100} (\bibinfo{year}{2021}), \bibinfo{pages}{106910}.
\newblock


\bibitem[Chen et~al\mbox{.}(2014)]%
        {2014modeling}
\bibfield{author}{\bibinfo{person}{Jun Chen}, \bibinfo{person}{Chaokun Wang}, {and} \bibinfo{person}{Jianmin Wang}.} \bibinfo{year}{2014}\natexlab{}.
\newblock \showarticletitle{Modeling the Interest-Forgetting Curve for Music Recommendation}. In \bibinfo{booktitle}{\emph{ACM International Conference on Multimedia (ICM)}}. \bibinfo{pages}{921--924}.
\newblock


\bibitem[Cheng et~al\mbox{.}(2016)]%
        {widedeep}
\bibfield{author}{\bibinfo{person}{Heng-Tze Cheng}, \bibinfo{person}{Levent Koc}, \bibinfo{person}{Jeremiah Harmsen}, \bibinfo{person}{Tal Shaked}, \bibinfo{person}{Tushar Chandra}, \bibinfo{person}{Hrishi Aradhye}, \bibinfo{person}{Glen Anderson}, \bibinfo{person}{Greg Corrado}, \bibinfo{person}{Wei Chai}, \bibinfo{person}{Mustafa Ispir}, {et~al\mbox{.}}} \bibinfo{year}{2016}\natexlab{}.
\newblock \showarticletitle{Wide \& Deep Learning for Recommender Systems}. In \bibinfo{booktitle}{\emph{Workshop on Deep Learning for Recommender Systems(DLRS)}}. \bibinfo{pages}{7--10}.
\newblock


\bibitem[Cheng et~al\mbox{.}(2020)]%
        {afn}
\bibfield{author}{\bibinfo{person}{Weiyu Cheng}, \bibinfo{person}{Yanyan Shen}, {and} \bibinfo{person}{Linpeng Huang}.} \bibinfo{year}{2020}\natexlab{}.
\newblock \showarticletitle{Adaptive Factorization Network: Learning Adaptive-Order Feature Interactions}. In \bibinfo{booktitle}{\emph{the AAAI Conference on Artificial Intelligence(AAAI)}}, Vol.~\bibinfo{volume}{34}. \bibinfo{pages}{3609--3616}.
\newblock


\bibitem[Chong et~al\mbox{.}(2023)]%
        {2023CT4Rec}
\bibfield{author}{\bibinfo{person}{Liu Chong}, \bibinfo{person}{Xiaoyang Liu}, \bibinfo{person}{Rongqin Zheng}, \bibinfo{person}{Lixin Zhang}, \bibinfo{person}{Xiaobo Liang}, \bibinfo{person}{Juntao Li}, \bibinfo{person}{Lijun Wu}, \bibinfo{person}{Min Zhang}, {and} \bibinfo{person}{Leyu Lin}.} \bibinfo{year}{2023}\natexlab{}.
\newblock \showarticletitle{CT4Rec: Simple yet Effective Consistency Training for Sequential Recommendation}. In \bibinfo{booktitle}{\emph{ACM SIGKDD Conference on Knowledge Discovery and Data Mining (KDD)}} (Long Beach, CA, USA). \bibinfo{pages}{3901–3913}.
\newblock
\showISBNx{9798400701030}


\bibitem[Concha-Carrasco et~al\mbox{.}(2023)]%
        {CONCHACARRASCO2023}
\bibfield{author}{\bibinfo{person}{José~A. Concha-Carrasco}, \bibinfo{person}{Miguel~A. Vega-Rodríguez}, {and} \bibinfo{person}{Carlos~J. Pérez}.} \bibinfo{year}{2023}\natexlab{}.
\newblock \showarticletitle{A Multi-Objective Artificial Bee Colony Approach for Profit-Aware Recommender Systems}.
\newblock \bibinfo{journal}{\emph{Information Sciences (IS)}}  \bibinfo{volume}{625} (\bibinfo{year}{2023}), \bibinfo{pages}{476--488}.
\newblock
\showISSN{0020-0255}


\bibitem[D{\'e}sid{\'e}ri(2012)]%
        {MGDA2012}
\bibfield{author}{\bibinfo{person}{Jean-Antoine D{\'e}sid{\'e}ri}.} \bibinfo{year}{2012}\natexlab{}.
\newblock \showarticletitle{Multiple-Gradient Descent Algorithm (MGDA) for Multiobjective Optimization}.
\newblock \bibinfo{journal}{\emph{Comptes Rendus Mathematique (CRM)}} \bibinfo{volume}{350}, \bibinfo{number}{5-6} (\bibinfo{year}{2012}), \bibinfo{pages}{313--318}.
\newblock


\bibitem[Ebbinghaus(1885)]%
        {1885memory}
\bibfield{author}{\bibinfo{person}{Hermann Ebbinghaus}.} \bibinfo{year}{1885}\natexlab{}.
\newblock \showarticletitle{Memory: A Contribution to Experimental Psychology. Teachers College, Columbia University, New York}.
\newblock \bibinfo{journal}{\emph{Trans. HA Ruger and CE Bussenius. Original work published}} (\bibinfo{year}{1885}).
\newblock


\bibitem[Faggioli et~al\mbox{.}(2020)]%
        {faggioli2020recency}
\bibfield{author}{\bibinfo{person}{Guglielmo Faggioli}, \bibinfo{person}{Mirko Polato}, {and} \bibinfo{person}{Fabio Aiolli}.} \bibinfo{year}{2020}\natexlab{}.
\newblock \showarticletitle{Recency Aware Collaborative Filtering for Next Basket Recommendation}. In \bibinfo{booktitle}{\emph{ACM Conference on User Modeling, Adaptation and Personalization (UMAP)}}. \bibinfo{pages}{80--87}.
\newblock


\bibitem[Ferreira et~al\mbox{.}(2023)]%
        {ferreiraforgetting}
\bibfield{author}{\bibinfo{person}{Jos{\'e}~{\^A}. Ferreira}, \bibinfo{person}{Edson~L. Valmorbida}, \bibinfo{person}{Bruno~G. Sato}, \bibinfo{person}{Bruno~P. Fuentes}, {and} \bibinfo{person}{Renan Botti}.} \bibinfo{year}{2023}\natexlab{}.
\newblock \showarticletitle{Forgetting Curve Models: A Systematic Review Aimed at Consolidating the Main Models and Outlining Possibilities for Future Research in Production}.
\newblock \bibinfo{journal}{\emph{Expert Systems (ES)}} (\bibinfo{year}{2023}), \bibinfo{pages}{e13405}.
\newblock


\bibitem[Fortes et~al\mbox{.}(2021)]%
        {2021inded}
\bibfield{author}{\bibinfo{person}{Reinaldo~S. Fortes}, \bibinfo{person}{Daniel~X. De~Sousa}, \bibinfo{person}{Dayanne~G. Coelho}, \bibinfo{person}{Anisio~M. Lacerda}, {and} \bibinfo{person}{Marcos~A. Gon{\c{c}}alves}.} \bibinfo{year}{2021}\natexlab{}.
\newblock \showarticletitle{Individualized Extreme Dominance (IndED): A New Preference-Based Method for Multi-Objective Recommender Systems}.
\newblock \bibinfo{journal}{\emph{Information Sciences (IS)}}  \bibinfo{volume}{572} (\bibinfo{year}{2021}), \bibinfo{pages}{558--573}.
\newblock


\bibitem[Frank et~al\mbox{.}(1956)]%
        {frankwolfe1956algorithm}
\bibfield{author}{\bibinfo{person}{Marguerite Frank}, \bibinfo{person}{Philip Wolfe}, {et~al\mbox{.}}} \bibinfo{year}{1956}\natexlab{}.
\newblock \showarticletitle{An Algorithm for Quadratic Programming}.
\newblock \bibinfo{journal}{\emph{Naval Research Logistics Quarterly(NRL)}} \bibinfo{volume}{3}, \bibinfo{number}{1-2} (\bibinfo{year}{1956}), \bibinfo{pages}{95--110}.
\newblock


\bibitem[Gao et~al\mbox{.}(2022)]%
        {gao2022kuairec}
\bibfield{author}{\bibinfo{person}{Chongming Gao}, \bibinfo{person}{Shijun Li}, \bibinfo{person}{Wenqiang Lei}, \bibinfo{person}{Jiawei Chen}, \bibinfo{person}{Biao Li}, \bibinfo{person}{Peng Jiang}, \bibinfo{person}{Xiangnan He}, \bibinfo{person}{Jiaxin Mao}, {and} \bibinfo{person}{Tat-Seng Chua}.} \bibinfo{year}{2022}\natexlab{}.
\newblock \showarticletitle{KuaiRec: A Fully-Observed Dataset and Insights for Evaluating Recommender Systems}. In \bibinfo{booktitle}{\emph{ACM International Conference on Information \& Knowledge Management(CIKM)}}. \bibinfo{pages}{540--550}.
\newblock


\bibitem[Gao et~al\mbox{.}(2019)]%
        {2019drcgr}
\bibfield{author}{\bibinfo{person}{Rong Gao}, \bibinfo{person}{Haifeng Xia}, \bibinfo{person}{Jing Li}, \bibinfo{person}{Donghua Liu}, \bibinfo{person}{Shuai Chen}, {and} \bibinfo{person}{Gang Chun}.} \bibinfo{year}{2019}\natexlab{}.
\newblock \showarticletitle{DRCGR: Deep Reinforcement Learning Framework Incorporating CNN and GAN-Based for Interactive Recommendation}. In \bibinfo{booktitle}{\emph{IEEE International Conference on Data Mining (ICDM)}}. IEEE, \bibinfo{pages}{1048--1053}.
\newblock


\bibitem[Ge et~al\mbox{.}(2022)]%
        {2022toward}
\bibfield{author}{\bibinfo{person}{Yingqiang Ge}, \bibinfo{person}{Xiaoting Zhao}, \bibinfo{person}{Lucia Yu}, \bibinfo{person}{Saurabh Paul}, \bibinfo{person}{Diane Hu}, \bibinfo{person}{Chu-Cheng Hsieh}, {and} \bibinfo{person}{Yongfeng Zhang}.} \bibinfo{year}{2022}\natexlab{}.
\newblock \showarticletitle{Toward Pareto Efficient Fairness-Utility Trade-Off in Recommendation Through Reinforcement Learning}. In \bibinfo{booktitle}{\emph{ACM International Conference on Web Search and Data Mining (WSDM)}}. \bibinfo{pages}{316--324}.
\newblock


\bibitem[Guo et~al\mbox{.}(2017)]%
        {deepfm}
\bibfield{author}{\bibinfo{person}{Huifeng Guo}, \bibinfo{person}{Ruiming Tang}, \bibinfo{person}{Yunming Ye}, \bibinfo{person}{Zhenguo Li}, {and} \bibinfo{person}{Xiuqiang He}.} \bibinfo{year}{2017}\natexlab{}.
\newblock \showarticletitle{DeepFM: A Factorization-Machine Based Neural Network for CTR Prediction}.
\newblock \bibinfo{journal}{\emph{International Joint Conference on Artificial Intelligence(IJCAI)}} (\bibinfo{year}{2017}).
\newblock


\bibitem[Hemmler et~al\mbox{.}(2023)]%
        {2023categorization}
\bibfield{author}{\bibinfo{person}{Yvonne~M. Hemmler}, \bibinfo{person}{Julian Rasch}, {and} \bibinfo{person}{Dirk Ifenthaler}.} \bibinfo{year}{2023}\natexlab{}.
\newblock \showarticletitle{A Categorization of Workplace Learning Goals for Multi-Stakeholder Recommender Systems: A Systematic Review}.
\newblock \bibinfo{journal}{\emph{TechTrends}} \bibinfo{volume}{67}, \bibinfo{number}{1} (\bibinfo{year}{2023}), \bibinfo{pages}{98--111}.
\newblock


\bibitem[Hofacker and Murphy(2009)]%
        {2009consumer}
\bibfield{author}{\bibinfo{person}{Charles~F. Hofacker} {and} \bibinfo{person}{Jamie Murphy}.} \bibinfo{year}{2009}\natexlab{}.
\newblock \showarticletitle{Consumer Web Page Search, Clicking Behavior and Reaction Time}.
\newblock \bibinfo{journal}{\emph{Direct Marketing: An International Journal}} \bibinfo{volume}{3}, \bibinfo{number}{2} (\bibinfo{year}{2009}), \bibinfo{pages}{88--96}.
\newblock


\bibitem[Huang et~al\mbox{.}(2023)]%
        {2023negative}
\bibfield{author}{\bibinfo{person}{Junjie Huang}, \bibinfo{person}{Ruobing Xie}, \bibinfo{person}{Qi Cao}, \bibinfo{person}{Huawei Shen}, \bibinfo{person}{Shaoliang Zhang}, \bibinfo{person}{Feng Xia}, {and} \bibinfo{person}{Xueqi Cheng}.} \bibinfo{year}{2023}\natexlab{}.
\newblock \showarticletitle{Negative Can Be Positive: Signed Graph Neural Networks for Recommendation}.
\newblock \bibinfo{journal}{\emph{Information Processing \& Management (IP\&M)}} \bibinfo{volume}{60}, \bibinfo{number}{4} (\bibinfo{year}{2023}), \bibinfo{pages}{103403}.
\newblock


\bibitem[Isufi et~al\mbox{.}(2021)]%
        {ISUFI2021}
\bibfield{author}{\bibinfo{person}{Elvin Isufi}, \bibinfo{person}{Matteo Pocchiari}, {and} \bibinfo{person}{Alan Hanjalic}.} \bibinfo{year}{2021}\natexlab{}.
\newblock \showarticletitle{Accuracy-Diversity Trade-Off in Recommender Systems via Graph Convolutions}.
\newblock \bibinfo{journal}{\emph{Information Processing \& Management (IP\&M)}} \bibinfo{volume}{58}, \bibinfo{number}{2} (\bibinfo{year}{2021}), \bibinfo{pages}{102459}.
\newblock
\showISSN{0306-4573}


\bibitem[Kingma and Ba(2014)]%
        {kingma2014adam}
\bibfield{author}{\bibinfo{person}{Diederik~P. Kingma} {and} \bibinfo{person}{Jimmy Ba}.} \bibinfo{year}{2014}\natexlab{}.
\newblock \showarticletitle{Adam: A Method for Stochastic Optimization}.
\newblock \bibinfo{journal}{\emph{International Conference on Learning Representations(ICLR)}} (\bibinfo{year}{2014}).
\newblock


\bibitem[Kuhn and Tucker(2013)]%
        {KKT}
\bibfield{author}{\bibinfo{person}{Harold~W. Kuhn} {and} \bibinfo{person}{Albert~W. Tucker}.} \bibinfo{year}{2013}\natexlab{}.
\newblock \showarticletitle{Nonlinear Programming}.
\newblock In \bibinfo{booktitle}{\emph{Traces and Emergence of Nonlinear Programming}}. \bibinfo{publisher}{Springer}, \bibinfo{pages}{247--258}.
\newblock


\bibitem[Kutlimuratov et~al\mbox{.}(2022)]%
        {2022modeling}
\bibfield{author}{\bibinfo{person}{Alpamis Kutlimuratov}, \bibinfo{person}{Akmalbek~B. Abdusalomov}, \bibinfo{person}{Rashid Oteniyazov}, \bibinfo{person}{Sanjar Mirzakhalilov}, {and} \bibinfo{person}{Taeg~K. Whangbo}.} \bibinfo{year}{2022}\natexlab{}.
\newblock \showarticletitle{Modeling and Applying Implicit Dormant Features for Recommendation via Clustering and Deep Factorization}.
\newblock \bibinfo{journal}{\emph{Sensors}} \bibinfo{volume}{22}, \bibinfo{number}{21} (\bibinfo{year}{2022}), \bibinfo{pages}{8224}.
\newblock


\bibitem[Li et~al\mbox{.}(2021)]%
        {2021multi}
\bibfield{author}{\bibinfo{person}{Hui Li}, \bibinfo{person}{Zhaoman Zhong}, \bibinfo{person}{Jun Shi}, \bibinfo{person}{Haining Li}, {and} \bibinfo{person}{Yong Zhang}.} \bibinfo{year}{2021}\natexlab{}.
\newblock \showarticletitle{Multi-Objective Optimization-Based Recommendation for Massive Online Learning Resources}.
\newblock \bibinfo{journal}{\emph{IEEE Sensors Journal(IEEE Sens. J.)}} \bibinfo{volume}{21}, \bibinfo{number}{22} (\bibinfo{year}{2021}), \bibinfo{pages}{25274--25281}.
\newblock


\bibitem[Li et~al\mbox{.}(2023)]%
        {2023relationship}
\bibfield{author}{\bibinfo{person}{Lei Li}, \bibinfo{person}{Yongfeng Zhang}, {and} \bibinfo{person}{Li Chen}.} \bibinfo{year}{2023}\natexlab{}.
\newblock \showarticletitle{On the Relationship between Explanation and Recommendation: Learning to Rank Explanations for Improved Performance}.
\newblock \bibinfo{journal}{\emph{ACM Transactions on Intelligent Systems and Technology (TIST)}} \bibinfo{volume}{14}, \bibinfo{number}{2} (\bibinfo{year}{2023}), \bibinfo{pages}{1--24}.
\newblock


\bibitem[Lin et~al\mbox{.}(2019)]%
        {2019Pareto}
\bibfield{author}{\bibinfo{person}{Xiao Lin}, \bibinfo{person}{Hongjie Chen}, \bibinfo{person}{Changhua Pei}, \bibinfo{person}{Fei Sun}, \bibinfo{person}{Xuanji Xiao}, \bibinfo{person}{Hanxiao Sun}, \bibinfo{person}{Yongfeng Zhang}, \bibinfo{person}{Wenwu Ou}, {and} \bibinfo{person}{Peng Jiang}.} \bibinfo{year}{2019}\natexlab{}.
\newblock \showarticletitle{A Pareto-Efficient Algorithm for Multiple Objective Optimization in E-Commerce Recommendation}. In \bibinfo{booktitle}{\emph{ACM Conference on Recommender Systems (RecSys)}}. \bibinfo{publisher}{{ACM}}, \bibinfo{pages}{20--28}.
\newblock


\bibitem[Lindauer et~al\mbox{.}(2022)]%
        {2022bayesianhyperparameter}
\bibfield{author}{\bibinfo{person}{Marius Lindauer}, \bibinfo{person}{Katharina Eggensperger}, \bibinfo{person}{Matthias Feurer}, \bibinfo{person}{Andr{\'e} Biedenkapp}, \bibinfo{person}{Difan Deng}, \bibinfo{person}{Carolin Benjamins}, \bibinfo{person}{Tim Ruhkopf}, \bibinfo{person}{Ren{\'e} Sass}, {and} \bibinfo{person}{Frank Hutter}.} \bibinfo{year}{2022}\natexlab{}.
\newblock \showarticletitle{SMAC3: A Versatile Bayesian Optimization Package for Hyperparameter Optimization}.
\newblock \bibinfo{journal}{\emph{The Journal of Machine Learning Research (JMLR)}} \bibinfo{volume}{23}, \bibinfo{number}{1} (\bibinfo{year}{2022}), \bibinfo{pages}{2475--2483}.
\newblock


\bibitem[Liu et~al\mbox{.}(2022)]%
        {2022self}
\bibfield{author}{\bibinfo{person}{Xiangbin Liu}, \bibinfo{person}{Shuqi Chen}, \bibinfo{person}{Liping Song}, \bibinfo{person}{Marcin Wo{\'z}niak}, {and} \bibinfo{person}{Shuai Liu}.} \bibinfo{year}{2022}\natexlab{}.
\newblock \showarticletitle{Self-Attention Negative Feedback Network for Real-Time Image Super-Resolution}.
\newblock \bibinfo{journal}{\emph{Journal of King Saud University-Computer and Information Sciences(J KING SAUD UNIV-COM)}} \bibinfo{volume}{34}, \bibinfo{number}{8} (\bibinfo{year}{2022}), \bibinfo{pages}{6179--6186}.
\newblock


\bibitem[L{\'o}pez-Ib{\'a}{\~n}ez et~al\mbox{.}(2016)]%
        {2016irace}
\bibfield{author}{\bibinfo{person}{Manuel L{\'o}pez-Ib{\'a}{\~n}ez}, \bibinfo{person}{J{\'e}r{\'e}mie Dubois-Lacoste}, \bibinfo{person}{Leslie~P. C{\'a}ceres}, \bibinfo{person}{Mauro Birattari}, {and} \bibinfo{person}{Thomas St{\"u}tzle}.} \bibinfo{year}{2016}\natexlab{}.
\newblock \showarticletitle{The Irace Package: Iterated Racing for Automatic Algorithm Configuration}.
\newblock \bibinfo{journal}{\emph{Operations Research Perspectives(Oper. Res. Perspect.)}}  \bibinfo{volume}{3} (\bibinfo{year}{2016}), \bibinfo{pages}{43--58}.
\newblock


\bibitem[Ma et~al\mbox{.}(2018b)]%
        {2018modeling}
\bibfield{author}{\bibinfo{person}{Jiaqi Ma}, \bibinfo{person}{Zhe Zhao}, \bibinfo{person}{Xinyang Yi}, \bibinfo{person}{Jilin Chen}, \bibinfo{person}{Lichan Hong}, {and} \bibinfo{person}{Ed~H. Chi}.} \bibinfo{year}{2018}\natexlab{b}.
\newblock \showarticletitle{Modeling Task Relationships in Multi-Task Learning with Multi-Gate Mixture-Of-Experts}. In \bibinfo{booktitle}{\emph{ACM SIGKDD International Conference on Knowledge Discovery \& Data Mining (KDD)}}. \bibinfo{pages}{1930--1939}.
\newblock


\bibitem[Ma et~al\mbox{.}(2018c)]%
        {2018MMOE}
\bibfield{author}{\bibinfo{person}{Jiaqi Ma}, \bibinfo{person}{Zhe Zhao}, \bibinfo{person}{Xinyang Yi}, \bibinfo{person}{Jilin Chen}, \bibinfo{person}{Lichan Hong}, {and} \bibinfo{person}{Ed~H Chi}.} \bibinfo{year}{2018}\natexlab{c}.
\newblock \showarticletitle{Modeling Task Relationships in Multi-task Learning with Multi-gate Mixture-of-Experts}. In \bibinfo{booktitle}{\emph{ACM SIGKDD International Conference on Knowledge Discovery \& Data Mining(KDD)}}. \bibinfo{pages}{1930--1939}.
\newblock


\bibitem[Ma et~al\mbox{.}(2020)]%
        {2020continuousPareto}
\bibfield{author}{\bibinfo{person}{Pingchuan Ma}, \bibinfo{person}{Tao Du}, {and} \bibinfo{person}{Wojciech Matusik}.} \bibinfo{year}{2020}\natexlab{}.
\newblock \showarticletitle{Efficient Continuous Pareto Exploration in Multi-Task Learning}. In \bibinfo{booktitle}{\emph{International Conference on Machine Learning, {ICML}}}. \bibinfo{publisher}{{PMLR}}, \bibinfo{pages}{6522--6531}.
\newblock


\bibitem[Ma et~al\mbox{.}(2018a)]%
        {2018ESMM}
\bibfield{author}{\bibinfo{person}{Xiao Ma}, \bibinfo{person}{Liqin Zhao}, \bibinfo{person}{Guan Huang}, \bibinfo{person}{Zhi Wang}, \bibinfo{person}{Zelin Hu}, \bibinfo{person}{Xiaoqiang Zhu}, {and} \bibinfo{person}{Kun Gai}.} \bibinfo{year}{2018}\natexlab{a}.
\newblock \showarticletitle{Entire Space Multi-Task Model: An Effective Approach for Estimating Post-Click Conversion Rate}. In \bibinfo{booktitle}{\emph{International ACM SIGIR Conference on Research \& Development in Information Retrieval(SIGIR)}}. \bibinfo{pages}{1137--1140}.
\newblock


\bibitem[Murphy et~al\mbox{.}(2006)]%
        {2006primacy}
\bibfield{author}{\bibinfo{person}{Jamie Murphy}, \bibinfo{person}{Charles Hofacker}, {and} \bibinfo{person}{Richard Mizerski}.} \bibinfo{year}{2006}\natexlab{}.
\newblock \showarticletitle{Primacy and Recency Effects on Clicking Behavior}.
\newblock \bibinfo{journal}{\emph{Journal of Computer-Mediated Communication (JCMC)}} \bibinfo{volume}{11}, \bibinfo{number}{2} (\bibinfo{year}{2006}), \bibinfo{pages}{522--535}.
\newblock


\bibitem[Neysiani et~al\mbox{.}(2019)]%
        {2019improve}
\bibfield{author}{\bibinfo{person}{Behzad~S. Neysiani}, \bibinfo{person}{Nasim Soltani}, \bibinfo{person}{Reza Mofidi}, {and} \bibinfo{person}{Mohammad~H. Nadimi-Shahraki}.} \bibinfo{year}{2019}\natexlab{}.
\newblock \showarticletitle{Improve Performance of Association Rule-Based Collaborative Filtering Recommendation Systems Using Genetic Algorithm}.
\newblock \bibinfo{journal}{\emph{International Journal of Information Technology and Computer Science (IGITCS)}} \bibinfo{volume}{11}, \bibinfo{number}{2} (\bibinfo{year}{2019}), \bibinfo{pages}{48--55}.
\newblock


\bibitem[Nitu et~al\mbox{.}(2021)]%
        {nitu2021improvising}
\bibfield{author}{\bibinfo{person}{Paromita Nitu}, \bibinfo{person}{Joseph Coelho}, {and} \bibinfo{person}{Praveen Madiraju}.} \bibinfo{year}{2021}\natexlab{}.
\newblock \showarticletitle{Improvising Personalized Travel Recommendation System with Recency Effects}.
\newblock \bibinfo{journal}{\emph{Big Data Mining and Analytics}} \bibinfo{volume}{4}, \bibinfo{number}{3} (\bibinfo{year}{2021}), \bibinfo{pages}{139--154}.
\newblock


\bibitem[Paparella(2022)]%
        {2022ParetoRS}
\bibfield{author}{\bibinfo{person}{Vincenzo Paparella}.} \bibinfo{year}{2022}\natexlab{}.
\newblock \showarticletitle{Pursuing Optimal Trade-Off Solutions in Multi-Objective Recommender Systems}. In \bibinfo{booktitle}{\emph{ACM Conference on Recommender Systems (RecSys)}}. \bibinfo{pages}{727–729}.
\newblock


\bibitem[Paparella et~al\mbox{.}(2023)]%
        {2023post-hoc}
\bibfield{author}{\bibinfo{person}{Vincenzo Paparella}, \bibinfo{person}{Vito~W. Anelli}, \bibinfo{person}{Franco~M. Nardini}, \bibinfo{person}{Raffaele Perego}, {and} \bibinfo{person}{Tommaso Di~Noia}.} \bibinfo{year}{2023}\natexlab{}.
\newblock \showarticletitle{Post-hoc Selection of Pareto-Optimal Solutions in Search and Recommendation}.
\newblock \bibinfo{journal}{\emph{ACM International Conference on Information and Knowledge Management(CIKM)}}  \bibinfo{volume}{abs/2306.12165} (\bibinfo{year}{2023}), \bibinfo{pages}{2013--2023}.
\newblock


\bibitem[Petchrompo et~al\mbox{.}(2022)]%
        {2022review}
\bibfield{author}{\bibinfo{person}{Sanyapong Petchrompo}, \bibinfo{person}{David~W. Coit}, \bibinfo{person}{Alexandra Brintrup}, \bibinfo{person}{Anupong Wannakrairot}, {and} \bibinfo{person}{Ajith~K. Parlikad}.} \bibinfo{year}{2022}\natexlab{}.
\newblock \showarticletitle{A Review of Pareto Pruning Methods for Multi-Objective Optimization}.
\newblock \bibinfo{journal}{\emph{Computers \& Industrial Engineering (CAIE)}}  \bibinfo{volume}{167} (\bibinfo{year}{2022}), \bibinfo{pages}{108022}.
\newblock


\bibitem[Ren et~al\mbox{.}(2023)]%
        {2023slate-aware}
\bibfield{author}{\bibinfo{person}{Yi Ren}, \bibinfo{person}{Xiao Han}, \bibinfo{person}{Xu Zhao}, \bibinfo{person}{Shenzheng Zhang}, {and} \bibinfo{person}{Yan Zhang}.} \bibinfo{year}{2023}\natexlab{}.
\newblock \showarticletitle{Slate-Aware Ranking for Recommendation}.
\newblock \bibinfo{journal}{\emph{ACM International Conference on Web Search and Data Mining (WSDM)}}, \bibinfo{pages}{499–507}.
\newblock
\showISBNx{9781450394079}


\bibitem[Ribeiro et~al\mbox{.}(2014)]%
        {2014POEA}
\bibfield{author}{\bibinfo{person}{Marco~T. Ribeiro}, \bibinfo{person}{Nivio Ziviani}, \bibinfo{person}{Edleno S.~D. Moura}, \bibinfo{person}{Itamar Hata}, \bibinfo{person}{Anisio Lacerda}, {and} \bibinfo{person}{Adriano Veloso}.} \bibinfo{year}{2014}\natexlab{}.
\newblock \showarticletitle{Multi-Objective Pareto-Efficient Approaches for Recommender Systems}.
\newblock \bibinfo{journal}{\emph{ACM Transactions on Intelligent Systems and Technology (TIST)}} \bibinfo{volume}{5}, \bibinfo{number}{4} (\bibinfo{year}{2014}), \bibinfo{pages}{1--20}.
\newblock


\bibitem[Sener and Koltun(2018)]%
        {sener2018multi}
\bibfield{author}{\bibinfo{person}{Ozan Sener} {and} \bibinfo{person}{Vladlen Koltun}.} \bibinfo{year}{2018}\natexlab{}.
\newblock \showarticletitle{Multi-Task Learning as Multi-Objective Optimization}.
\newblock \bibinfo{journal}{\emph{Advances in Neural Information Processing Systems(NeurIPS)}}  \bibinfo{volume}{31} (\bibinfo{year}{2018}).
\newblock


\bibitem[Shrivastava et~al\mbox{.}(2023)]%
        {2023deep}
\bibfield{author}{\bibinfo{person}{Rahul Shrivastava}, \bibinfo{person}{Dilip~S. Sisodia}, {and} \bibinfo{person}{Naresh~K. Nagwani}.} \bibinfo{year}{2023}\natexlab{}.
\newblock \showarticletitle{Deep Neural Network-Based Multi-Stakeholder Recommendation System Exploiting Multi-Criteria Ratings for Preference Learning}.
\newblock \bibinfo{journal}{\emph{Expert Systems with Applications (ESWA)}}  \bibinfo{volume}{213} (\bibinfo{year}{2023}), \bibinfo{pages}{119071}.
\newblock


\bibitem[Sun et~al\mbox{.}(2023)]%
        {2023berd+}
\bibfield{author}{\bibinfo{person}{Yatong Sun}, \bibinfo{person}{Xiaochun Yang}, \bibinfo{person}{Zhu Sun}, {and} \bibinfo{person}{Bin Wang}.} \bibinfo{year}{2023}\natexlab{}.
\newblock \showarticletitle{BERD+: A Generic Sequential Recommendation Framework by Eliminating Unreliable Data with Item-and Attribute-Level Signals}.
\newblock \bibinfo{journal}{\emph{ACM Transactions on Information Systems (TOIS)}} (\bibinfo{year}{2023}).
\newblock


\bibitem[Tang et~al\mbox{.}(2020)]%
        {2020progressive}
\bibfield{author}{\bibinfo{person}{Hongyan Tang}, \bibinfo{person}{Junning Liu}, \bibinfo{person}{Ming Zhao}, {and} \bibinfo{person}{Xudong Gong}.} \bibinfo{year}{2020}\natexlab{}.
\newblock \showarticletitle{Progressive Layered Extraction (PLE): A Novel Multi-Task Learning (MTL) Model for Personalized Recommendations}. In \bibinfo{booktitle}{\emph{ACM Conference on Recommender Systems (RecSys)}}. \bibinfo{pages}{269--278}.
\newblock


\bibitem[Vyas et~al\mbox{.}(2021)]%
        {2021drivebfr}
\bibfield{author}{\bibinfo{person}{Jayant Vyas}, \bibinfo{person}{Debasis Das}, {and} \bibinfo{person}{Santanu Chaudhury}.} \bibinfo{year}{2021}\natexlab{}.
\newblock \showarticletitle{DriveBFR: Driver Behavior and Fuel-Efficiency-Based Recommendation System}.
\newblock \bibinfo{journal}{\emph{IEEE Transactions on Computational Social Systems (TCSS)}} \bibinfo{volume}{9}, \bibinfo{number}{5} (\bibinfo{year}{2021}), \bibinfo{pages}{1446--1455}.
\newblock


\bibitem[Wang et~al\mbox{.}(2021)]%
        {2021attribute}
\bibfield{author}{\bibinfo{person}{Suhua Wang}, \bibinfo{person}{Lisa Zhang}, \bibinfo{person}{Mengying Yu}, \bibinfo{person}{Yuling Wang}, \bibinfo{person}{Zhiqiang Ma}, {and} \bibinfo{person}{Yu Zhao}.} \bibinfo{year}{2021}\natexlab{}.
\newblock \showarticletitle{Attribute-Aware Multi-Task Recommendation}.
\newblock \bibinfo{journal}{\emph{The Journal of Supercomputing(J Supercomput)}}  \bibinfo{volume}{77} (\bibinfo{year}{2021}), \bibinfo{pages}{4419--4437}.
\newblock


\bibitem[Wang and Cao(2021)]%
        {wang2021interactive}
\bibfield{author}{\bibinfo{person}{Wei Wang} {and} \bibinfo{person}{Longbing Cao}.} \bibinfo{year}{2021}\natexlab{}.
\newblock \showarticletitle{Interactive Sequential Basket Recommendation by Learning Basket Couplings and Positive/Negative Feedback}.
\newblock \bibinfo{journal}{\emph{ACM Transactions on Information Systems (TOIS)}} \bibinfo{volume}{39}, \bibinfo{number}{3} (\bibinfo{year}{2021}), \bibinfo{pages}{1--26}.
\newblock


\bibitem[Wang et~al\mbox{.}(2023a)]%
        {2023learning}
\bibfield{author}{\bibinfo{person}{Yueqi Wang}, \bibinfo{person}{Yoni Halpern}, \bibinfo{person}{Shuo Chang}, \bibinfo{person}{Jingchen Feng}, \bibinfo{person}{Elaine~Y. Le}, \bibinfo{person}{Longfei Li}, \bibinfo{person}{Xujian Liang}, \bibinfo{person}{Min-Cheng Huang}, \bibinfo{person}{Shane Li}, \bibinfo{person}{Alex Beutel}, {et~al\mbox{.}}} \bibinfo{year}{2023}\natexlab{a}.
\newblock \showarticletitle{Learning from Negative User Feedback and Measuring Responsiveness for Sequential Recommenders}. In \bibinfo{booktitle}{\emph{ACM Conference on Recommender Systems (RecSys)}}. \bibinfo{pages}{1049--1053}.
\newblock


\bibitem[Wang et~al\mbox{.}(2023b)]%
        {2023multi}
\bibfield{author}{\bibinfo{person}{Yuhao Wang}, \bibinfo{person}{Ha~T. Lam}, \bibinfo{person}{Yi Wong}, \bibinfo{person}{Ziru Liu}, \bibinfo{person}{Xiangyu Zhao}, \bibinfo{person}{Yichao Wang}, \bibinfo{person}{Bo Chen}, \bibinfo{person}{Huifeng Guo}, {and} \bibinfo{person}{Ruiming Tang}.} \bibinfo{year}{2023}\natexlab{b}.
\newblock \showarticletitle{Multi-Task Deep Recommender Systems: A Survey}.
\newblock \bibinfo{journal}{\emph{CoRR}}  \bibinfo{volume}{abs/2302.03525} (\bibinfo{year}{2023}).
\newblock


\bibitem[Wu et~al\mbox{.}(2020)]%
        {wu2020neural}
\bibfield{author}{\bibinfo{person}{Chuhan Wu}, \bibinfo{person}{Fangzhao Wu}, \bibinfo{person}{Yongfeng Huang}, {and} \bibinfo{person}{Xing Xie}.} \bibinfo{year}{2020}\natexlab{}.
\newblock \showarticletitle{Neural News Recommendation with Negative Feedback}.
\newblock \bibinfo{journal}{\emph{CCF Transactions on Pervasive Computing and Interaction (CCF TPCI)}}  \bibinfo{volume}{2} (\bibinfo{year}{2020}), \bibinfo{pages}{178--188}.
\newblock


\bibitem[Wu et~al\mbox{.}(2023)]%
        {20231hyperparameter}
\bibfield{author}{\bibinfo{person}{Di Wu}, \bibinfo{person}{Bo Sun}, {and} \bibinfo{person}{Mingsheng Shang}.} \bibinfo{year}{2023}\natexlab{}.
\newblock \showarticletitle{Hyperparameter Learning for Deep Learning-Based Recommender Systems}.
\newblock \bibinfo{journal}{\emph{IEEE Transactions on Services Computing (TSC)}} (\bibinfo{year}{2023}), \bibinfo{pages}{2699--2712}.
\newblock


\bibitem[Wu et~al\mbox{.}(2022b)]%
        {2022multistakeholderPareto}
\bibfield{author}{\bibinfo{person}{Haolun Wu}, \bibinfo{person}{Chen Ma}, \bibinfo{person}{Bhaskar Mitra}, \bibinfo{person}{Fernando Diaz}, {and} \bibinfo{person}{Xue Liu}.} \bibinfo{year}{2022}\natexlab{b}.
\newblock \showarticletitle{A Multi-Objective Optimization Framework for Multi-Stakeholder Fairness-Aware Recommendation}.
\newblock \bibinfo{journal}{\emph{ACM Transactions on Information Systems (TOIS)}} \bibinfo{volume}{41}, \bibinfo{number}{2} (\bibinfo{year}{2022}), \bibinfo{pages}{1--29}.
\newblock


\bibitem[Wu et~al\mbox{.}(2022a)]%
        {RS2022survey}
\bibfield{author}{\bibinfo{person}{Le Wu}, \bibinfo{person}{Xiangnan He}, \bibinfo{person}{Xiang Wang}, \bibinfo{person}{Kun Zhang}, {and} \bibinfo{person}{Meng Wang}.} \bibinfo{year}{2022}\natexlab{a}.
\newblock \showarticletitle{A Survey on Accuracy-Oriented Neural Recommendation: From Collaborative Filtering to Information-Rich Recommendation}.
\newblock \bibinfo{journal}{\emph{IEEE Transactions on Knowledge and Data Engineering (TKDE)}} \bibinfo{volume}{35}, \bibinfo{number}{5} (\bibinfo{year}{2022}), \bibinfo{pages}{4425--4445}.
\newblock


\bibitem[Wu et~al\mbox{.}(2010)]%
        {2010LambdaMART}
\bibfield{author}{\bibinfo{person}{Qiang Wu}, \bibinfo{person}{Christopher J.~C. Burges}, \bibinfo{person}{Krysta~M. Svore}, {and} \bibinfo{person}{Jianfeng Gao}.} \bibinfo{year}{2010}\natexlab{}.
\newblock \showarticletitle{Adapting Boosting for Information Retrieval Measures}.
\newblock \bibinfo{journal}{\emph{Information Retrieval}}  \bibinfo{volume}{13} (\bibinfo{year}{2010}), \bibinfo{pages}{254--270}.
\newblock


\bibitem[Xi et~al\mbox{.}(2023)]%
        {2023device}
\bibfield{author}{\bibinfo{person}{Yunjia Xi}, \bibinfo{person}{Weiwen Liu}, \bibinfo{person}{Yang Wang}, \bibinfo{person}{Ruiming Tang}, \bibinfo{person}{Weinan Zhang}, \bibinfo{person}{Yue Zhu}, \bibinfo{person}{Rui Zhang}, {and} \bibinfo{person}{Yong Yu}.} \bibinfo{year}{2023}\natexlab{}.
\newblock \showarticletitle{On-device Integrated Re-ranking with Heterogeneous Behavior Modeling}. In \bibinfo{booktitle}{\emph{ACM SIGKDD Conference on Knowledge Discovery and Data Mining (KDD)}}. \bibinfo{pages}{5225--5236}.
\newblock


\bibitem[Xie et~al\mbox{.}(2021)]%
        {2021paretoQM}
\bibfield{author}{\bibinfo{person}{Ruobing Xie}, \bibinfo{person}{Yanlei Liu}, \bibinfo{person}{Shaoliang Zhang}, \bibinfo{person}{Rui Wang}, \bibinfo{person}{Feng Xia}, {and} \bibinfo{person}{Leyu Lin}.} \bibinfo{year}{2021}\natexlab{}.
\newblock \showarticletitle{Personalized Approximate Pareto-Efficient Recommendation}. In \bibinfo{booktitle}{\emph{ACM Web Conference (WWW)}} (Ljubljana, Slovenia). \bibinfo{pages}{3839–3849}.
\newblock
\showISBNx{9781450383127}


\bibitem[Xu(2013)]%
        {2013social}
\bibfield{author}{\bibinfo{person}{Qian Xu}.} \bibinfo{year}{2013}\natexlab{}.
\newblock \showarticletitle{Social Recommendation, Source Credibility, and Recency: Effects of News Cues in a Social Bookmarking Website}.
\newblock \bibinfo{journal}{\emph{Journalism \& Mass Communication Quarterly (JMCQ)}} \bibinfo{volume}{90}, \bibinfo{number}{4} (\bibinfo{year}{2013}), \bibinfo{pages}{757--775}.
\newblock


\bibitem[Yannam et~al\mbox{.}(2023a)]%
        {2023enhancing}
\bibfield{author}{\bibinfo{person}{V~R. Yannam}, \bibinfo{person}{Jitendra Kumar}, \bibinfo{person}{Korra~S. Babu}, {and} \bibinfo{person}{Bidyut~K. Patra}.} \bibinfo{year}{2023}\natexlab{a}.
\newblock \showarticletitle{Enhancing the Accuracy of Group Recommendation Using Slope One}.
\newblock \bibinfo{journal}{\emph{The Journal of Supercomputing(J Supercomput)}} \bibinfo{volume}{79}, \bibinfo{number}{1} (\bibinfo{year}{2023}), \bibinfo{pages}{499--540}.
\newblock


\bibitem[Yannam et~al\mbox{.}(2023b)]%
        {2023improving}
\bibfield{author}{\bibinfo{person}{V~R. Yannam}, \bibinfo{person}{Jitendra Kumar}, \bibinfo{person}{Korra~S. Babu}, {and} \bibinfo{person}{Bibhudatta Sahoo}.} \bibinfo{year}{2023}\natexlab{b}.
\newblock \showarticletitle{Improving Group Recommendation Using Deep Collaborative Filtering Approach}.
\newblock \bibinfo{journal}{\emph{International Journal of Information Technology (IJOIT)}} \bibinfo{volume}{15}, \bibinfo{number}{3} (\bibinfo{year}{2023}), \bibinfo{pages}{1489--1497}.
\newblock


\bibitem[Yin et~al\mbox{.}(2022)]%
        {2022building}
\bibfield{author}{\bibinfo{person}{Fulian Yin}, \bibinfo{person}{Yanyan Pan}, \bibinfo{person}{Pei Su}, {and} \bibinfo{person}{Yanyan Wang}.} \bibinfo{year}{2022}\natexlab{}.
\newblock \showarticletitle{Building User Interest Model for TV Recommendation with Label-Based Memory Forgetting-Enhancement Model}.
\newblock \bibinfo{journal}{\emph{Multimedia Tools and Applications (MTAP)}} \bibinfo{volume}{81}, \bibinfo{number}{18} (\bibinfo{year}{2022}), \bibinfo{pages}{26307--26330}.
\newblock


\bibitem[Yu et~al\mbox{.}(2021)]%
        {2021selective}
\bibfield{author}{\bibinfo{person}{Xu Yu}, \bibinfo{person}{Qinglong Peng}, \bibinfo{person}{Lingwei Xu}, \bibinfo{person}{Feng Jiang}, \bibinfo{person}{Junwei Du}, {and} \bibinfo{person}{Dunwei Gong}.} \bibinfo{year}{2021}\natexlab{}.
\newblock \showarticletitle{A Selective Ensemble Learning Based Two-Sided Cross-Domain Collaborative Filtering Algorithm}.
\newblock \bibinfo{journal}{\emph{Information Processing \& Management (IP\&M)(Inf Process Manag)}} \bibinfo{volume}{58}, \bibinfo{number}{6} (\bibinfo{year}{2021}), \bibinfo{pages}{102691}.
\newblock


\bibitem[Zaizi et~al\mbox{.}(2023)]%
        {2023multi-objectivesurvey}
\bibfield{author}{\bibinfo{person}{Fatima~E. Zaizi}, \bibinfo{person}{Sara Qassimi}, {and} \bibinfo{person}{Said Rakrak}.} \bibinfo{year}{2023}\natexlab{}.
\newblock \showarticletitle{Multi-Objective Optimization with Recommender Systems: {A} Systematic Review}.
\newblock \bibinfo{journal}{\emph{Information Systems}}  \bibinfo{volume}{117} (\bibinfo{year}{2023}), \bibinfo{pages}{102233}.
\newblock


\bibitem[Zhang et~al\mbox{.}(2019)]%
        {2019indexed}
\bibfield{author}{\bibinfo{person}{Lei Zhang}, \bibinfo{person}{Shangshang Yang}, \bibinfo{person}{Xinpeng Wu}, \bibinfo{person}{Fan Cheng}, \bibinfo{person}{Ying Xie}, {and} \bibinfo{person}{Zhiting Lin}.} \bibinfo{year}{2019}\natexlab{}.
\newblock \showarticletitle{An Indexed Set Representation Based Multi-Objective Evolutionary Approach for Mining Diversified Top-K High Utility Patterns}.
\newblock \bibinfo{journal}{\emph{Engineering Applications of Artificial Intelligence (EAAI)}}  \bibinfo{volume}{77} (\bibinfo{year}{2019}), \bibinfo{pages}{9--20}.
\newblock


\bibitem[Zhao et~al\mbox{.}(2023)]%
        {2023copr}
\bibfield{author}{\bibinfo{person}{Zhishan Zhao}, \bibinfo{person}{Jingyue Gao}, \bibinfo{person}{Yu Zhang}, \bibinfo{person}{Shuguang Han}, \bibinfo{person}{Siyuan Lou}, \bibinfo{person}{Xiang-Rong Sheng}, \bibinfo{person}{Zhe Wang}, \bibinfo{person}{Han Zhu}, \bibinfo{person}{Yuning Jiang}, \bibinfo{person}{Jian Xu}, {et~al\mbox{.}}} \bibinfo{year}{2023}\natexlab{}.
\newblock \showarticletitle{COPR: Consistency-Oriented Pre-Ranking for Online Advertising}.
\newblock \bibinfo{journal}{\emph{ACM International Conference on Information and Knowledge Management(CIKM)}}  \bibinfo{volume}{abs/2306.03516} (\bibinfo{year}{2023}), \bibinfo{pages}{4974–4980}.
\newblock


\bibitem[Zheng et~al\mbox{.}(2023)]%
        {2023tourism}
\bibfield{author}{\bibinfo{person}{Xiaoyao Zheng}, \bibinfo{person}{Baoting Han}, {and} \bibinfo{person}{Zhen Ni}.} \bibinfo{year}{2023}\natexlab{}.
\newblock \showarticletitle{Tourism Route Recommendation Based on A Multi-Objective Evolutionary Algorithm Using Two-Stage Decomposition and Pareto Layering}.
\newblock \bibinfo{journal}{\emph{IEEE/CAA Journal of Automatica Sinica(JAS)}} \bibinfo{volume}{10}, \bibinfo{number}{2} (\bibinfo{year}{2023}), \bibinfo{pages}{486--500}.
\newblock


\bibitem[Zhou et~al\mbox{.}(2020)]%
        {zhou2020s3}
\bibfield{author}{\bibinfo{person}{Kun Zhou}, \bibinfo{person}{Hui Wang}, \bibinfo{person}{Wayne~Xin Zhao}, \bibinfo{person}{Yutao Zhu}, \bibinfo{person}{Sirui Wang}, \bibinfo{person}{Fuzheng Zhang}, \bibinfo{person}{Zhongyuan Wang}, {and} \bibinfo{person}{Ji-Rong Wen}.} \bibinfo{year}{2020}\natexlab{}.
\newblock \showarticletitle{S3-rec: Self-supervised learning for sequential recommendation with mutual information maximization}. In \bibinfo{booktitle}{\emph{ACM International Conference on Information \& Knowledge Management (CIKM)}}. \bibinfo{pages}{1893--1902}.
\newblock


\bibitem[Zhu et~al\mbox{.}(2023)]%
        {2023integrated}
\bibfield{author}{\bibinfo{person}{Menghui Zhu}, \bibinfo{person}{Wei Xia}, \bibinfo{person}{Weiwen Liu}, \bibinfo{person}{Yifan Liu}, \bibinfo{person}{Ruiming Tang}, {and} \bibinfo{person}{Weinan Zhang}.} \bibinfo{year}{2023}\natexlab{}.
\newblock \showarticletitle{Integrated Ranking for News Feed with Reinforcement Learning}. In \bibinfo{booktitle}{\emph{ACM Web Conference(WWW)}}. \bibinfo{pages}{480--484}.
\newblock


\bibitem[Zou et~al\mbox{.}(2021)]%
        {2021MOQM}
\bibfield{author}{\bibinfo{person}{Feng Zou}, \bibinfo{person}{Debao Chen}, \bibinfo{person}{Qingzheng Xu}, \bibinfo{person}{Ziqi Jiang}, {and} \bibinfo{person}{Jiahui Kang}.} \bibinfo{year}{2021}\natexlab{}.
\newblock \showarticletitle{A Two-Stage Personalized Recommendation Based on Multi-Objective Teaching–Learning-Based Optimization with Decomposition}.
\newblock \bibinfo{journal}{\emph{Neurocomputing}}  \bibinfo{volume}{452} (\bibinfo{year}{2021}), \bibinfo{pages}{716--727}.
\newblock
\showISSN{0925-2312}


\end{thebibliography}

\end{document}